\documentclass[a4]{aa}

\usepackage{epsf}
\usepackage{txfonts}
\usepackage{color}

\def\changed{}

\def\ion#1#2{#1\,{\sc #2}}

\def\HeI{\ion{He}{i}}
\def\HeII{\ion{He}{ii}}

\def\NII{\ion{N}{ii}}
\def\NIII{\ion{N}{iii}}

\def\NV{\ion{N}{v}}

\def\CIII{\ion{C}{iii}}
\def\CIV{\ion{C}{iv}}

\def\DM{$D\!M$}

\begin{document}

\title{The Galactic WN stars}
\subtitle{\changed Spectral analyses with line-blanketed model atmospheres 
versus stellar evolution models with and without rotation}

\author{W.-R. Hamann \and G. Gr\"afener \and A. Liermann} 

\offprints{W.-R.\ Hamann \\
\email{wrh@astro.physik.uni-potsdam.de}}

\institute{Lehrstuhl Astrophysik der Universit\"at Potsdam,
Am Neuen Palais 10, D-14469 Potsdam, Germany}

\date{Received \today; accepted ...}

\abstract
{Very massive stars pass through the Wolf-Rayet (WR)
stage before they finally explode. Details of their evolution have not
yet been safely established, and their physics are not well understood. Their
spectral analysis requires adequate model atmospheres, which have been
developed step by step during the past decades and account in their
recent version for line blanketing by the millions of lines from iron
and iron-group elements. However, only very few WN stars have been
re-analyzed by means of line-blanketed models yet.}
{The quantitative spectral analysis of a large sample of Galactic WN 
stars with the most advanced generation of model atmospheres 
should provide an empirical basis for various studies about the origin,
evolution, and physics of the Wolf-Rayet stars and their powerful 
winds.}
{We analyze a large sample of Galactic WN stars by
means of the Potsdam Wolf-Rayet (PoWR) model atmospheres, which account
for iron line blanketing and clumping. The results are
compared with a synthetic population, generated from the
Geneva tracks for massive star evolution.} 
{We obtain a homogeneous set of
stellar and atmospheric parameters for the Galactic WN stars, 
partly revising earlier results.} 
{Comparing the results of our spectral analyses
of the Galactic WN stars with the predictions of the Geneva evolutionary
calculations, we conclude that there is rough qualitative agreement.
However, the quantitative discrepancies are still severe, and there is
no preference for the tracks that account for the effects of rotation.
It seems that the evolution of massive stars is still not satisfactorily
understood.}


\maketitle

\section{Introduction}

Very massive stars pass through the Wolf-Rayet (WR)
stage before they finally explode as supernovae or, possibly,
$\gamma$-ray bursts. The WR stars are important sources of ionizing photons,
momentum, and chemical elements. However, the evolution of massive stars
has not yet been safely established, and their physics is not well understood.

The empirical knowledge about WR stars is hampered by difficulties
with their spectral analysis. Adequate model atmospheres, which account
for the non-LTE physics and their supersonic expansion, have been
developed step by step during the past decades. In a previous paper we
presented a comprehensive analysis of the Galactic WN stars from
their helium, hydrogen, and nitrogen spectra (Hamann \& Koesterke
\cite{HK98}, hereafter quoted as Paper\,I). The major deficiency of this
generation of model atmospheres was its neglect of line blanketing by
the millions of lines from iron and iron-group elements. Hillier \&
Miller (\cite{HiMi98}) were the first to  include this important effect
in their code {\sc cmfgen}. The line-blanketed version of the Potsdam 
Wolf-Rayet (PoWR) model atmosphere code became
available with Gr\"afener et al.\ (\cite{blanketing02}).

The improved models provide a much better fit to the observed spectra,
and hence lead to more reliable determination of the stellar
parameters. For many WN stars, a substantial revision of previous
results can be expected. However, only very few WN stars have been
re-analyzed so far by means of line-blanketed models (Herald et al.\
\cite{Herald01}; Marchenko et al.\ \cite{Marchenko+al2004}). In the
present paper we now re-analyze all Galactic WN stars for which 
observed spectra are available to us.

The best way to analyze a larger sample of stars systematically is
first to establish ``model grids'', i.e.\ sets of model atmospheres
and  synthetic spectra that cover the relevant range of parameters.
For WN stars we have already prepared such grids of iron-line blanketed 
models (Hamann \& Gr\"afener \cite{blanket-WN}), which will be used in
the present paper. 

The main objective of analyzing the Galactic WN stars is to understand the
evolution of massive stars. The empirical results from our previous
comprehensive analyses were in conflict with the evolutionary
calculations. However, {\changed with the application of the more
advanced models, we will revise the empirical stellar parameters.
Moreover, the evolutionary models have also been significantly improved
recently, since the effects of stellar rotation are now taken into
account (Meynet \& Maeder \cite{MM03}). Hence, the question of whether stellar
evolution theory still conflicts with the empirical stellar
parameters should be reassessed. For this purpose, we use these new
evolutionary tracks for generating synthetic stellar populations and
compare them with the results from our analyses of the Galactic WN
sample.}

The paper is organized as follows. In Sect.\,2 we briefly
characterize the applied model atmospheres. The program stars and
observational data are introduced in Sect.\,3. The spectral analyses and
their results are given in Sect.\,4, together with a long list of
detailed comments on individual WN stars. Our empirical results are
compared with evolutionary models in Sect.\,5, especially by means of a
population synthesis.

\section{The models}

The Potsdam Wolf-Rayet (PoWR) model atmospheres (see Hamann \&
Gr\"afener 2004, and references therein, for more details) are based on
the ``standard'' assumptions (spherically symmetric, stationary
mass-loss). The velocity field is pre-specified in the standard way. For
the supersonic part we adopt the usual $\beta$-law, with the terminal
velocity $\varv_\infty$ being a free parameter. The exponent $\beta$ is set
to unity throughout this work. In the subsonic region, the velocity field
is defined such that a hydrostatic density stratification is approached.

The ``stellar radius'' $R_*$, which is the inner boundary of our model
atmosphere, corresponds by definition to a Rosseland optical depth of
20. The ``stellar temperature'' $T_*$ is defined by the luminosity $L$
and the stellar radius $R_*$ via the Stefan-Boltzmann law; i.e.\ $T_*$
denotes the effective temperature referring to the radius $R_*$.

Wind inhomogeneities (``clumping'') are now accounted for in a
first-order approximation, assuming that optically thin clumps fill a
volume fraction $f_{\rm V}$ while the interclump space is void. Thus the
matter density in the clumps is higher by a factor $D = f_{\rm V}^{-1}$,
compared to an un-clumped model with the same parameters. $D$\,=\,4 is assumed
throughout this paper. The Doppler velocity $\varv_{\rm D}$, which reflects
random motions on small scales (``microturbulence''), is set to $100 \,
{\rm km \, s^{-1}}$.

The major improvement of the models compared to those applied in
Paper\,I is the inclusion of line blanketing by iron and other
iron-group elements. About $10^5$ energy levels and $10^7$ line
transitions between those levels are taken into account in the
approximation of the ``superlevel'' approach.

The PoWR code was used to establish two grids of WN-type models
(Hamann \& Gr\"afener \cite{blanket-WN}). One grid is for hydrogen-free
stars, the other one contains 20\% of hydrogen (per mass). Internet
access to the synthetic spectra is provided to the
community\footnote{\tt http://www.astro.physik.uni-potsdam.de/PoWR.html}.

Complex model ions of H, He, \NIII\ -- \NV\, and \CIII\ -- \CIV\ are
taken into account in our models. The total number of Non-LTE levels is 261, 
including 72 iron superlevels. Due to the neglect of \NII, the
corresponding lines that might be detectable in the coolest WNL stars
are not included in our synthetic spectra. Because of the unsettled 
questions about a proper treatment of dielectronic recombination, this 
process is not taken into account. 

The analyses described in the present paper mainly rely on the PoWR 
grid models. In addition, individual models were calculated for some 
stars with special settings for their hydrogen abundance and 
terminal wind velocity.

\section{Program stars, observational data}
\label{sect:programstars}

The VIIth Catalogue of Galactic Wolf-Rayet Stars (van der Hucht
\cite{vdH}, hereafter ``the WR Catalog'') lists
227 stars in total. Among them, 127 stars\footnote{Actually one star
less, as we believe that WR\,109 (classified in the WR catalog as
WN5h+?) is erroneously counted here as a WN star. The typical brightness
of a WN star would imply an implausible large distance
(\DM\,=\,17.4\,mag), placing it beyond our Galaxy. From the binary period
and the optical flickering, it has been concluded already (Steiner et
al.\ \cite{Steiner+al1988}, \cite{Steiner+al1999}; Steiner \& Diaz
\cite{Steiner+Diaz1998}) that WR\,109 is a cataclysmic variable (CV).}
belong to the WN sequence considered here, while the others belong to
the WC sequence or are of intermediate spectral type (WN/WC).

Many of the WN stars in the catalog are highly reddened and therefore
invisible or at least very faint in visual light. Our study started with
74 WN stars for which we had sufficient spectral observations at our
disposal. As we restrict the present analyses to single-star spectra, 
well-established binaries with composed spectra (such as
WR\,139 alias V444\,Cyg) have been omitted from the beginning. 

But so far our sample still comprised many stars that have been
suspected of binarity, on more or less sound basis. We considered all
these cases in detail, checking the literature and inspecting our
spectral fits for indications of binarity (cf.\ Hamann \& Gr\"afener
\cite{H+G-canada}). For 11 objects we finally conclude that they are
indeed binaries, where the companion of the WN stars contributes more
than 15\% of the total light in the visual. Those stars were excluded from the
present single-star analyses. They are marked in
Table\,\ref{table:parameters} as ``composite spectrum'', the reasoning
given below in the ``Comments on individual stars''
(Sect.\,\ref{sect:individual}).

Among the remaining 63 stars (see Table\,\ref{table:parameters}), which
we are going to analyze in the present paper, there are still four stars
that are most likely close binaries, but the light contamination from
the non-WN companion can be neglected. These stars are marked by the
superscript $^b$ at their WR number in Table\,\ref{table:parameters}, 
column\,(1), indicating their possible origin from close-binary evolution.
Note that six stars of our present sample (WR\,28, WR\,63, WR\,71, 
WR\,85, WR\,94, WR\,107) have not been analyzed before. 

The observational material is mostly the same as used in our previous papers 
and as was published in our atlas of WN spectra (see Hamann et al.\
\cite{wn-atlas} for more details). These spectra were taken at the
``Deutsch-Spanisches Astronomisches Zentrum (DSAZ)'', Calar Alto, Spain,
and at the European Southern Observatory (ESO), La Silla, Chile.
For four stars (WR\,6, WR\,22, WR\,24, WR\,78)
we can employ high-resolution optical spectra from the ESO archive 
obtained with UVES. All optical spectra are not flux-calibrated and 
therefore were normalized to the continuum ``by eye'' before being 
compared with normalized synthetic spectra. 

Additionally, we retrieved IUE low-resolution spectra from the
archive for all program stars when available. As these spectra are 
flux-calibrated, we can fully exploit this information. We divide
the flux-calibrated observation by the (reddened) model continuum and
thus obtain the comparison plot with the normalized line spectrum of the
model (cf.\ Fig.\,\ref{fig:wr061-lines}, upper panel).

For fitting the spectral energy distribution, we use the IUE flux and
narrow-band visual photometry ($\varv$, $b$ bands). In addition we
employ the homogeneous set of infrared photometry ($J$, $H$, $K$ bands)
from the 2MASS survey that is available now. This extension of the
wavelength range turns out to be very useful for a precise determination
of the interstellar absorption, especially for highly reddened stars 
without UV observation.   

The spectral types given in Table\,\ref{table:parameters}, column\,(2),
are taken basically from the WR catalog.  For WNE stars we append
\mbox{``-w''} or \mbox{``-s''} to the classification, indicating whether
the emission lines are weak or strong, respectively. The criterion is
whether the equivalent width of the \HeII\ line at 5412\,\AA\ is smaller
or larger than 37\,\AA\, as introduced in earlier papers.
 
In the older classification schemes, there was a clear dichotomy between
early and late WN subtypes (WNE and WNL, respectively). The early
subtypes (subtype numbers up to WN6) generally showed no hydrogen,
while the late subtypes (WN7 and higher) comprised the stars with rather
high hydrogen abundance. With the revised classification scheme used in
the WR catalog, this dichotomy is unfortunately mixed up. Now the WN6
subtype also includes a few stars that are typical members of the WNL
group (namely WR\,24, WR\,25, and WR\,85, as evident from their high
hydrogen abundance and their whole spectra appearance, cf.\ the atlas of
WR spectra from Hamann et al.\ \cite{wn-atlas}). Vice versa, a few stars
classified as WN7 or WN8 actually belong to the WNE subclass (namely
WR\,55, WR\,63, WR\,74, WR\,84, WR\,91, WR\,100, WR\,120, and WR\,123). 
We indicate these crosswise memberships to the WNL or WNE class in
parentheses behind the spectral subtype in column\,(2) of
Table\,\ref{table:parameters}. When discussing mean parameters of the
subclasses, we assign these stars accordingly.

\section{The analyses} 

\subsection{The line spectrum}
\label{sect:linespectrum}

\begin{figure*}[!tb]
\centering
\epsfxsize=0.8\textwidth
\mbox{\epsffile{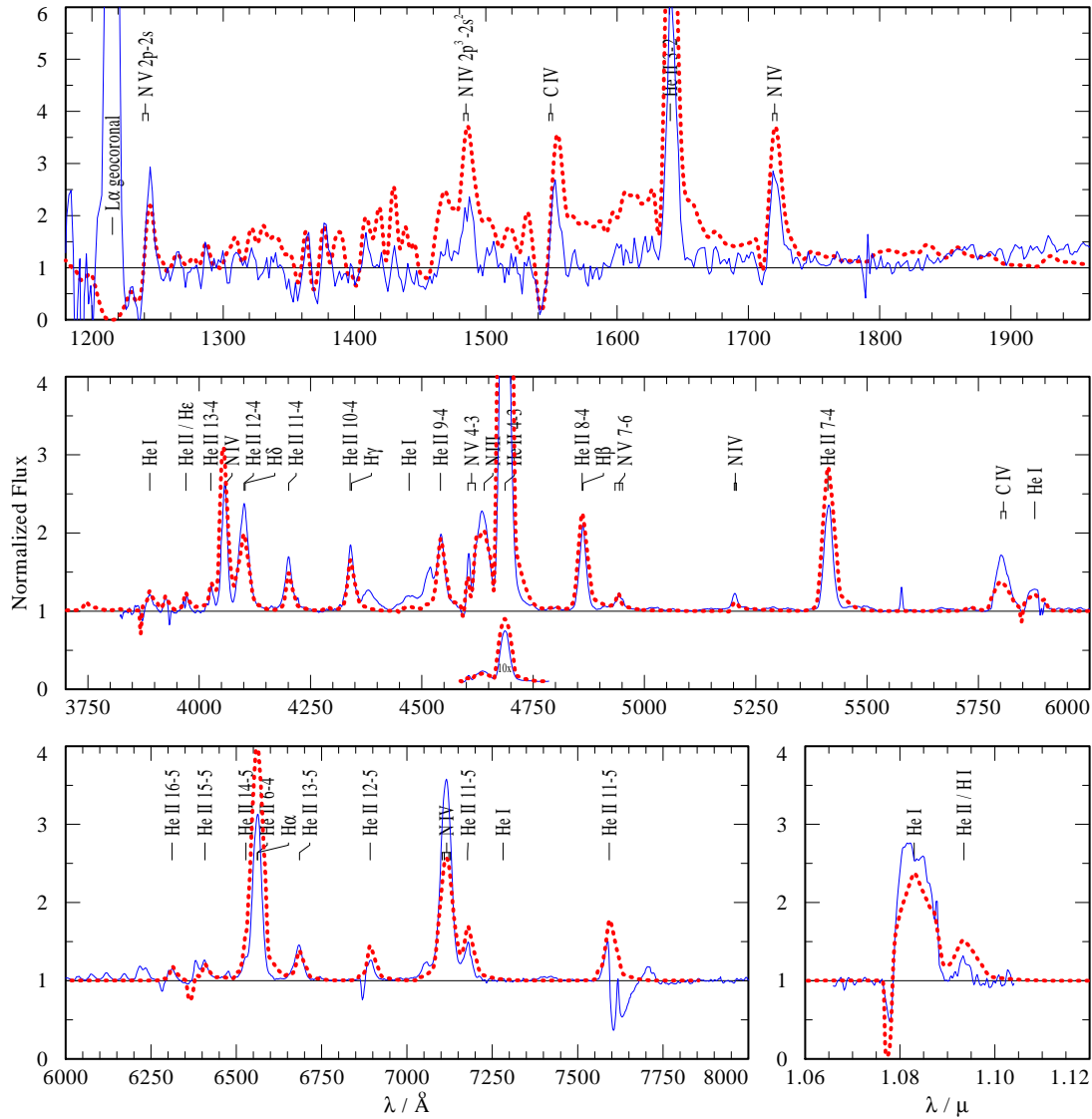}}
\caption{Spectrum of WR\,61 (solid line), together with the WNE grid
model 09-14 ($T_\ast$ = 63\,kK, $\log (R_{\rm t}/R_\odot)$ = 0.7, dotted
line). The IUE spectrum (top panel) has been divided by the reddened
model continuum (see Fig.\,\ref{fig:wr061-sed}), while the optical and
the near-IR spectra were normalized by eye.}
\label{fig:wr061-lines}
\end{figure*}

As has been described in many of our previous papers, the spectroscopic 
parameters of a Wolf-Rayet type stellar atmosphere are the stellar 
temperature $T_\ast$ and a second parameter that is related to the 
mass-loss rate and that we define in the form of the ``transformed 
radius'' 

\begin{equation}
R_{\rm t} = R_* \left[\frac{\varv_\infty}{2500 \, {\rm km}\,{\rm s^{-1}}} 
\left/
\frac{\dot{M}\ \sqrt{D} }
{10^{-4} \, M_\odot \, {\rm yr^{-1}}}\right]^{2/3} \right. 
\, .
\label{eq:rt}
\end{equation}

For a given fixed chemical composition and a given stellar temperature
$T_*$, synthetic spectra from Wolf-Rayet model atmospheres of different
mass-loss rates, stellar radii, and terminal wind velocities yield almost
the same emission-line equivalent widths, if they agree in their
``transformed radius'' $R_{\rm t}$. Note that the
``spectroscopic parameters'' $T_\ast$ and $R_{\rm t}$ do not involve 
the stellar distance.

Therefore the model grids (Hamann \& Gr\"afener \cite{blanket-WN}) are 
calculated with stepwise varied $\log T_\ast$ and $\log R_{\rm t}$, while 
all other parameters are kept fixed. Two grids have been established, 
one for WNE stars (without hydrogen, $\varv_\infty$ = 1600\,km/s) and 
one for WNL stars (20\% hydrogen, $\varv_\infty$ = 1000\,km/s). For the 
analyses, we now plot the (normalized) observed spectrum for all stars, 
together with a grid model, selecting the best-fitting one. For a couple
of stars we calculate individual models with suitably adapted parameters
(hydrogen abundance, terminal wind speed). {\changed A typical line fit
is shown in Fig.\,\ref{fig:wr061-lines}. We provide
the fits for our whole sample as {\em online material}.}

\begin{figure*}[!tb]
\centering
\epsfxsize=0.75\textwidth
\mbox{\epsffile{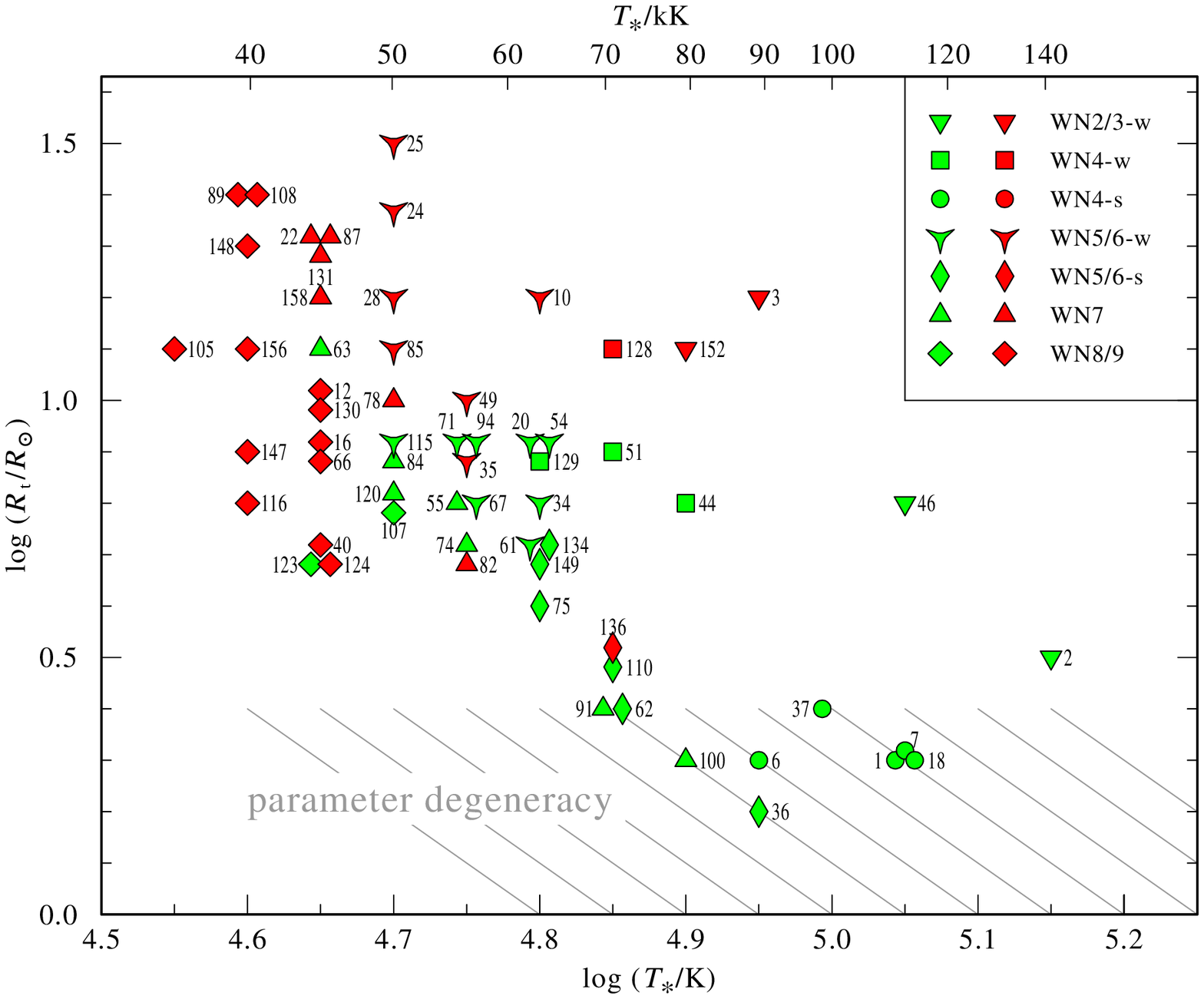}}
\caption{Parameters of the Galactic WN stars (labels: WR-Numbers). Shown
is the ``transformed radius'' $R_{\rm t}$ (see text) versus the stellar
temperature $T_\ast$. Dark/red filled symbols refer to stars with detectable
hydrogen, while light/green filled symbols denote hydrogen-free stars. The
symbol shapes are coding for the spectral subtype (see inlet). In the
lower, hatched part of the diagram the winds are very thick, and the
parameter space is degenerate such that any stars located in this part
can be arbitrarily shifted parallel to the grey lines.}
\label{fig:rtt}
\end{figure*}

The parameters $T_\ast$ and $R_{\rm t}$ of the best-fitting model are
tabulated in Table\,\ref{table:parameters}, columns (3) and (4) for all
program stars. Plotting these parameters in Fig.\,\ref{fig:rtt} reveals
that the cooler, late-type WN stars generally have thin winds and show
atmospheric hydrogen, while the hotter, early-type stars (WN2 ... WN6)
have thicker winds and are mostly hydrogen-free.

{\changed The WNL stars populate a temperature strip between 40 and
50\,kK. In comparison to Paper\,I, the WNL stars became about 10\,kK
hotter. The WNE stars scatter over a wide range of stellar temperatures
(50 ... 140\,kK). These temperatures have also been revised upwards in
many cases compared to Paper\,I. The higher stellar temperatures are a
consequence of the line-blanketing effects that are now taken into
account in the models. The $R_{\rm t}$ values are not much different
from Paper\,I. Note that these results are not systematically affected
by the introduction of clumping, as the clumping parameter is included
in the definition of the transformed radius (Eq.\,(\ref{eq:rt})), and
models with same $R_{\rm t}$ show very similar spectra.}

A few early-type WN stars with strong lines (suffix -s) fall into the
domain of parameter degeneracy, where the whole emergent spectrum is
formed in rapidly expanding layers. For stars with $\log (R_{\rm 
t}/R_\odot) < 0.4$, the spectra depend mainly on the product
$R_{\rm t}T_\ast^2$, but hardly on the individual values of these two 
parameters. Hence the location of those stars in the $R_{\rm 
t}$-$T_\ast$-plane can be shifted arbitrarily along the grey lines in 
Fig.\,\ref{fig:rtt} without spoiling the line fit. Note that along these 
lines the mass-loss rate is constant (in a model grid with fixed 
luminosity). The reason for this degeneracy is 
that spherically expanding atmospheres with $\varv(r) = \mathrm{constant}$ 
and the same ratio $L/\dot{M}^{4/3}$ are homologues.  

The terminal wind velocities (Table\,\ref{table:parameters},
column\,(5)) are taken from Paper\,I in most cases. For a number of
stars we revised $\varv_\infty$ after closer inspection of our line
fits. For those stars that were analyzed in detail in the meantime, we
adopt $\varv_\infty$ from these sources: \\
WR\,108: Crowther et al.\ (\cite{CrowtherI}); \\
WR\,22, WR\,24, WR\,25, WR\,78, 
WR\,120, WR\,123, WR\,124, WR\,156: Crowther et al.\ (\cite{CrowtherII});\\
WR\,128, WR\,152: Crowther et al.\ (\cite{CrowtherIV}); \\ 
WR\,16, WR\,40: Herald et al.\ (\cite{Herald01}). 

The hydrogen abundance, given in Table\,\ref{table:parameters}
column\,(6) in terms of mass fractions $X_{\rm H}$, is also taken from
Paper\,I or from the detailed analyses quoted in the previous paragraph.
For three stars that are new in the sample, $X_{\rm H}$ is estimated
for the first time (WR\,28, WR\,63, WR\,85). If hydrogen abundances have
been revised with respect to Paper\,I, this is mentioned in each
individual case in Sect.\,\ref{sect:individual}.

\subsection{The spectral energy distribution}
\label{sect:sed}
\begin{figure*}[!tb]
\centering
\epsfxsize=0.8\textwidth
\mbox{\epsffile{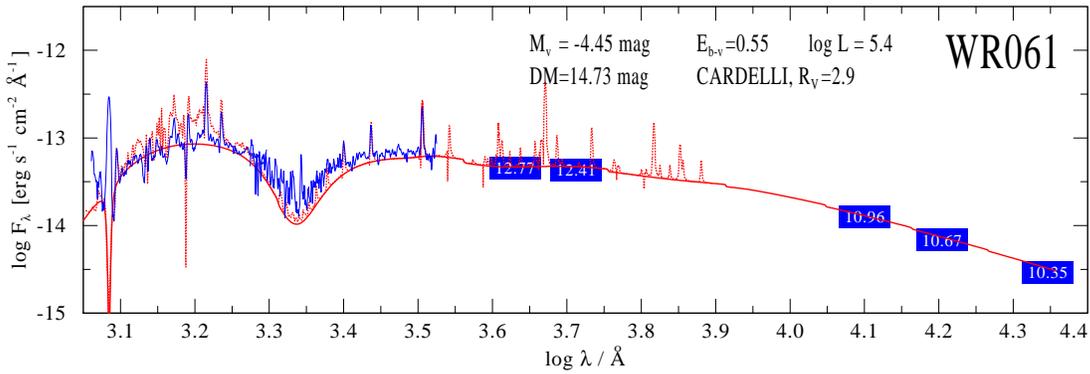}}
\caption{Spectral energy distribution (SED). The IUE observation (noisy
line) and photometry in $b$, $\varv$, $J$, $H$ and $K$ (blocks labelled
with magnitude) is compared with the reddened flux (smooth line: model
continuum; dotted line: model spectrum with lines) from the WNE grid
model 09-14 ($T_\ast$ = 63\,kK, $\log (R_{\rm t}/R_\odot)$ = 0.7). 
Reddening parameters 
and $\log L$ are suitably adjusted. The adopted absolute magnitude of
$M_\varv$ = -4.45\,mag for the WN5 subtype implies a distance modulus
of \DM = 14.73\,mag.
}
\label{fig:wr061-sed}
\end{figure*}

Now we exploit the information provided by the spectral energy
distribution (SED). We fit the whole SED from the UV (if
IUE observations are available) to infrared wavelengths. The IR photometry
($J$, $H$, $K$) can be found for all program stars in the 2MASS catalog.
This information turns out to be extremely useful, especially for highly
reddened objects. Visual photometry is taken from the WR catalog. Note
that the monochromatic magnitudes $b$, $\varv$ as defined by Smith
(\cite{Smith1968}), are used for WR photometry. The color excess in this
system is related to the $E_{B-V}$ in the Johnson system by $E_{B-V} =
1.21 E_{b-\varv}$.

The flux from the model (as selected by the line fit described in the
previous subsection) is plotted together with the observation. An
example is shown in Fig.\,\ref{fig:wr061-sed}, {\changed while
corresponding plots for the whole sample of stars are included in the
{\em online material}}. In order to achieve a fit, $E_{b-\varv}$ is
suitably adjusted by hand (Table\,\ref{table:parameters}, column\,(7)).
By default, the reddening law from Seaton (\cite{Seaton1979}) is applied.
It is augmented by interstellar L$\alpha$ absorption in the UV and by
the data from Moneti et al.\ (\cite{Moneti+al2001}) in the IR. In cases
where Seaton's law does not allow us to reproduce the observed SED
satisfactorily, we try to improve the fit with the help of anomalous
reddening. The reddening laws from Cardelli et al.\
(\cite{Cardelli+al1989}) and from Fitzpartick (\cite{Fitzpatrick1999})
provide the adjustable parameter $R_V = A_V/E_{B-V}$, while in Seaton's
law $R_V = 3.1$ is fixed. Our optimum choice for the reddening law and
$R_V$ (if applicable) is given in Table\,\ref{table:parameters},
column\,(8). Similar anomalous reddening often concerns a whole group of
neighboring stars, e.g.\ the Car\,OB1 members WR\,22, WR\,24, and WR\,25.

The chosen reddening implies extinction at the $\varv$ band ($A_\varv
= E_{b-\varv} ( 1.21 R_V + 0.36)$, cf.\ Lundstroem \& Stenholm
\cite{LS1984}). Hence, when the distance modulus \DM\ of the object is
known (see below for the discussion of distances), the absolute visual
magnitude $M_\varv$ follows from the observed (apparent) magnitude
$\varv$ via $M_\varv = \varv - D\!M - A_\varv$. Vice versa, if \DM\ is
not known, it follows from an adopted absolute visual magnitude
$M_\varv$.

The model flux is diluted according to the distance modulus \DM\ and
reddened according to the adopted reddening law and parameter(s). At
the first attempt, the model flux will not match the observed flux level.
Thanks to the (approximate) scalability of the models (while the
transformed radius is kept constant), the luminosity can be suitably
adjusted (remember that all grid models are calculated for $\log
L/L_\odot$ = 5.3).

The obtained stellar luminosity, radius, and mass-loss rate thus depend
on the adopted stellar distance. Fortunately, about one third of the
program stars are members of open clusters or associations. Those stars
are identified in Table\,\ref{table:parameters} by an arrow that points
from column\,(9) to the right. For a few stars their unclear membership
is discussed individually in Sect.\,\ref{sect:individual}.

The distance moduli for the cluster/association members
(Table\,\ref{table:parameters}, column\,(9)) are taken from the WR
catalog. Compared to Paper\,I, some of these distances have been
revised significantly (as also discussed in Sect.\,\ref{sect:individual}
for the individual stars). The new values are not always more plausible
than the older ones, but their discussion is beyond the scope
of this paper. The resulting stellar parameters can be easily
scaled when different distances are assumed.

\begin{figure}[!tb]
\centering
\epsfxsize=\columnwidth
\mbox{\epsffile{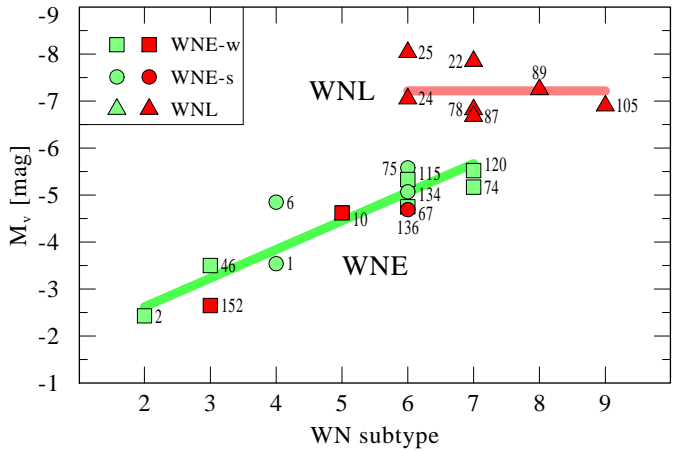}}
\caption{Absolute visual magnitudes of Galactic WN stars with known
distance. The dark (red) symbols refer to stars with detectable hydrogen, 
while symbols with light (green) filling stand for hydrogen-free stars.
The thick lines indicate the relations which we adopted for stars of 
unknown distance.}
\label{fig:mvcalib}
\end{figure}

\begin{table}
\caption{Absolute magnitudes of the different WN subtypes, as adopted for 
stars of unknown distance
\label{table:mvcalib}}
\begin{center}
\begin{tabular}{lc} 
\hline\hline 
Subtype & $M_\mathrm v$ [mag] \\ 
\hline 
\emph{hydrogen-free (WNE)} & \\
 WN2   &  -2.62 \\
 WN3   &  -3.23 \\
 WN4   &  -3.84 \\
 WN5   &  -4.45 \\
 WN6   &  -5.06 \\
 WN7   &  -5.67 \\
\emph{with hydrogen (WNL)} & \\
 WN6-9 &  -7.22 \\
\hline 
\end{tabular}
\end{center}
\end{table}

The absolute visual magnitude $M_\varv$ shows a nice correlation with
the WN subtype (Fig.\,\ref{fig:mvcalib}). However, it is important to
distinguish between hydrogen-rich and hydrogen-poor stars because the WNE
and WNL subclasses overlap in the WN6 and WN7 subtype according to the
current classification scheme, as discussed in
Sect.\,\ref{sect:programstars}. All WNL stars seem to have a similar
brightness, while for the WNE there is roughly a linear relation between
$M_\varv$ and subtype number. On this basis we adopt a ``typical'' absolute
magnitude for each subtype, as given in Table\,\ref{table:mvcalib}, and
apply this calibration for those stars for which the distance is not
known. For these stars, the arrow between columns\,(9) and (10) in
Table\,\ref{table:parameters} points to the left, i.e.\ from $M_\varv$
to \DM\ corresponding to the flow of information.

{\changed Of course, the ``typical'' brightness for the WNL stars must
be taken with care. The three stars in the Car\,OB1 association WR\,22,
WR\,24, and WR\,25 might be especially bright; at least, their hydrogen
abundance is outstandingly large. Without these three stars, the mean 
$M_\varv$ of the remaining four WNL stars would be -6.91\,mag, i.e. 0.31\,mag 
fainter than adopted here (Table\,\ref{table:mvcalib}).}

\begin{figure}[!tb]
\centering
\epsfxsize=\columnwidth
\mbox{\epsffile{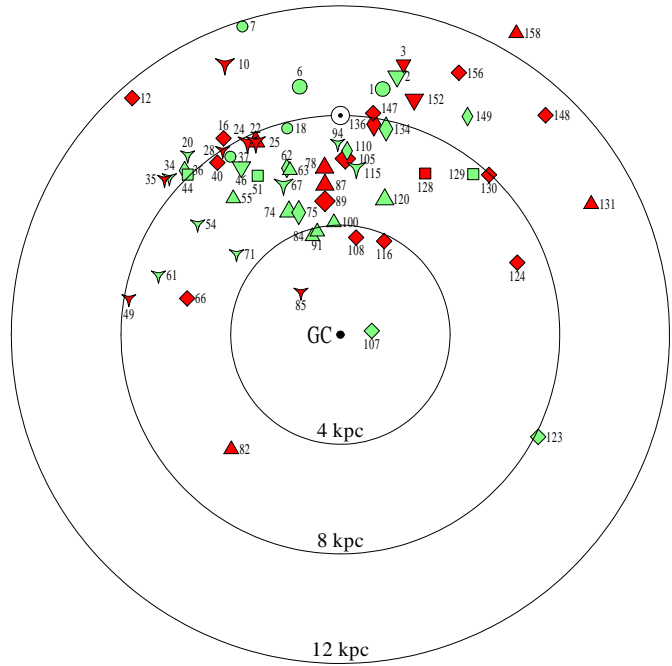}}
\caption{Galactic position of the analyzed WN stars. The meaning of the
symbols is the same as in Fig.\,\ref{fig:rtt}. Stars with distances
known from cluster/association membership are represented by bigger
symbols, while smaller symbols rely on our $M_\varv$ calibration of
the WN subtypes. The sun ($\odot$) and the Galactic Center (``GC'') are
indicated.}
\label{fig:gal-KOS}
\end{figure}

In Fig.\,\ref{fig:gal-KOS} we plot the position of our program stars in 
the Galactic plane, taking the distance and the Galactic longitude (and 
neglecting the Galactic latitude, which is small). The location of 
the WN stars along the nearby spiral arms looks plausible; at least 
there is no obvious outlier to indicate that its distance 
modulus is drastically wrong.   

Now with the distance modulus fixed for all program stars, we can 
establish the SED fits and determine the appropriate scaling factor 
of the model flux. Thus we obtain the stellar luminosity $L$ and, by means 
of Eq.\,(1), the mass-loss rate $\dot{M}$ and the stellar radius $R_\ast$ 
as given in columns\,(11--13) of Table\,\ref{table:parameters}. 

{\changed Compared to the results of Paper\,I, the luminosities obtained
now are significantly larger, especially for the WNL stars. This is only
partly due to the $M_\varv$ calibration discussed above. More important,
this is a consequence of the higher stellar temperatures that we derive
now with the line-blanketed model atmospheres. The mass-loss rates became
lower than in Paper\,I, simply because of the adopted clumping with $D$
= 4.}


\def\DM{$D\!M$}

\def\WNEs{\mbox{WNE-s}}
\def\WNEw{\mbox{WNE-w}}
\def\*{$^*$}

\def\r{\,\raisebox{0.4ex}{\tiny$\rightarrow$}\,}
\def\l{\,\raisebox{0.4ex}{\tiny$\leftarrow$}\,}
\def\noobs{\multicolumn{3}{c}{-- no observation --}}
\def\comp{\multicolumn{3}{c}{-- composite spectrum --}}
\def\s{\rule[0mm]{0mm}{4.5mm}}
\def\a{$^a$}
\def\b{$^b$}
\tabcolsep 4.3pt
\begin{table*}[!hbtp]
\caption[]{
Parameters of the Galactic single WN stars
\label{table:parameters}
} 
\small
\begin{flushleft}
\begin{tabular}{ l l r r r r r l r c r l r r }
\hline\hline 
\s WR & 
Spectral subtype &
$T_*$ & 
$\log R_{\rm t}$ & 
\multicolumn{1}{c}{$\varv_\infty$} & 
$X_{\rm H}$ & 
$E_{b-\varv}$ & 
Law\a  & 
\DM\ \hspace{5mm} $M_\varv$ &
$R_*$& log\,$\dot{M}^\dagger$ & 
log\,$L$ & 
$\frac{{\dot M} \varv_{\infty}}{L/c}$ & 
$M$ \\

&      
& 
[kK] & 
$[R_{\odot}]$ & 
\multicolumn{1}{c}{[km/s]} &
[\%] &
[mag]& 
$R_V$  & 
[mag]\ \ \ [mag] & 
$[R_{\odot}]$ & 
[$M_{\odot}$/yr] &
[$L_{\odot}$] &
&
[$M_{\odot}$] \\

   (1) & (2)           &(3)      & (4)   & \multicolumn{1}{c}{(5)}    
                                             &  (6) & (7)&  (8) & (9) \hspace{0.6cm} (10)
                                                                                   & (11) & (12) & (13)& (14) &(15) \\
\hline 
\s 1    & WN4-s         & 112.2 & 0.3 & 1900 &  0 & 0.67 & S     & 11.3 \r -3.54 & 1.33 & -4.7 & 5.4  &  7.7 & 15 \\
   2    & WN2-w         & 141.3 & 0.5 & 1800 &  0 & 0.44 & C 3.0 & 12.0 \r -2.43 & 0.89 & -5.3 & 5.45 &  1.7 & 16 \\
   3    & WN3h-w        &  89.1 & 1.2 & 2700 & 20 & 0.35 & C 3.4 & 12.4 \l -3.23 & 2.65 & -5.4 & 5.6  &  1.2 & 19 \\
   6    & WN4-s         &  89.1 & 0.3 & 1700 &  0 & 0.12 & S     & 11.3 \r -4.85 & 2.65 & -4.3 & 5.6  & 11.0 & 19 \\
   7    & WN4-s         & 112.2 & 0.3 & 1600 &  0 & 0.53 & S     & 13.4 \l -3.84 & 1.41 & -4.7 & 5.45 &  5.3 & 16 \\
   10   & WN5ha-w       &  63.1 & 1.2 & 1100 & 25 & 0.58 & C 3.1 & 13.3 \r -4.62 & 5.61 & -5.3 & 5.65 &  0.6 & 20 \\
   12\b & WN8h + ?      &  44.7 & 1.0 & 1200 & 27 & 0.80 & C 3.7 & 14.4 \l -7.22 & 21.1 & -4.1 & 6.2  &  2.7 & 44 \\
   16   & WN8h          &  44.7 & 0.9 &  650 & 25 & 0.55 & C 3.4&  13.2 \l -7.22 & 19.9 & -4.3 & 6.15 &  1.2 & 41 \\
   18   & WN4-s         & 112.2 & 0.3 & 1800 &  0 & 0.75 & C 3.6 & 11.5 \l -3.84 & 1.49 & -4.6 & 5.5  &  6.5 & 17 \\
   20   & WN5-w         &  63.1 & 0.9 & 1200 &  0 & 1.28 & S     & 13.8 \l -4.45 & 5.29 & -4.7 & 5.6  &  5.4 & 19 \\
 
\s 21   & WN5 + O4-6    &  \comp \\
   22\b & WN7h + O9III-V&  44.7 & 1.3 & 1785 & 44 & 0.35 & C 3.8 & 12.55\r -7.85 & 31.5 & -4.2 & 6.55 &  1.8 & 74 \\
   24   & WN6ha-w (WNL) &  50.1 & 1.35& 2160 & 44 & 0.24 & C 3.1 & 12.55\r -7.05 & 19.9 & -4.4 & 6.35 &  1.7 & 54 \\
   25\b & WN6h-w + ? (WNL)&50.1 & 1.5 & 2480 & 53 & 0.63 & C 4.5 & 12.55\r -8.04 & 33.4 & -4.3 & 6.8  &  1.0 & 110\\
   28   & WN6(h)-w      &  50.1 & 1.2 & 1200 & 20 & 1.20 & S     & 13.1 \l -5.06 & 8.89 & -5.0 & 5.65 &  1.3 & 20 \\
   31   & WN4 + O8V     &  \comp \\
   34   & WN5-w         &  63.1 & 0.8 & 1400 &  0 & 1.18 & S     & 14.1 \l -4.45 & 4.72 & -4.7 & 5.5  &  4.0 & 17 \\
   35   & WN6h-w        &  56.2 & 0.9 & 1100 & 22 & 1.15 & S     & 14.2 \l -5.06 & 6.67 & -4.8 & 5.6  &  2.3 & 19 \\
   36   & WN5-s         &  89.1 & 0.2 & 1900 &  0 & 1.00 & S     & 13.9 \l -4.45 & 1.88 & -4.3 & 5.3  & 23.0 & 13 \\
   37   & WN4-s         & 100.0 & 0.4 & 2150 &  0 & 1.63 & S     & 13.2 \l -3.84 & 1.88 & -4.6 & 5.5  &  9.3 & 17 \\

\s 40   & WN8h          &  44.7 & 0.7 & 650  & 23 & 0.40 & C 3.4 & 13.3 \l -7.22 & 17.7 & -4.1 & 6.05 &  2.5 & 35 \\
   44   & WN4-w         &  79.4 & 0.8 & 1400 &  0 & 0.62 & C 3.6 & 13.9 \l -3.84 & 3.15 & -5.0 & 5.55 &  1.9 & 18 \\
   46   & WN3p-w        & 112.2 & 0.8 & 2300 &  0 & 0.30 & F 3.6 & 13.05\r -3.50 & 2.11 & -5.1 & 5.8  &  1.6 & 25 \\
   47   & WN6 + O5V     &  \comp \\
   49   & WN5(h)-w      &  56.2 & 1.0 & 1450 & 25 & 0.80 & S     & 15.0 \l -4.45 & 5.29 & -5.0 & 5.4  &  3.2 & 15 \\
   51   & WN4-w         &  70.8 & 0.9 & 1500 &  0 & 1.40 & S     & 12.9 \l -3.84 & 3.75 & -5.0 & 5.5  &  2.3 & 17 \\
   54   & WN5-w         &  63.1 & 0.9 & 1500 &  0 & 0.82 & S     & 14.1 \l -4.45 & 5.29 & -4.8 & 5.6  &  3.0 & 19 \\
   55   & WN7 (WNE-w)   &  56.2 & 0.8 & 1200 &  0 & 0.65 & C 3.6 & 13.5 \l -5.67 & 8.39 & -4.4 & 5.8  &  3.5 & 25 \\
   61   & WN5-w         &  63.1 & 0.7 & 1400 &  0 & 0.55 & C 2.9 & 14.7 \l -4.45 & 4.21 & -4.7 & 5.4  &  5.9 & 15 \\
   62   & WN6-s         &  70.8 & 0.4 & 1800 &  0 & 1.73 & S     & 12.2 \l -5.06 & 3.54 & -4.2 & 5.45 & 19.0 & 16 \\

\s 63   & WN7 (WNE-w)   &  44.7 & 1.1 & 1700 &  0 & 1.54 & C 3.1 & 12.2 \l -5.67 & 11.2 & -4.6 & 5.65 &  5.3 & 20 \\
   66   & WN8(h)        &  44.7 & 0.9 & 1500 &  5 & 1.00 & S     & 14.8 \l -7.22 & 19.9 & -3.9 & 6.15 &  6.2 & 41 \\
   67   & WN6-w         &  56.2 & 0.8 & 1500 &  0 & 1.05 & S     & 12.6 \r -4.75 & 5.29 & -4.6 & 5.4  &  6.8 & 15 \\
   71   & WN6-w         &  56.2 & 0.9 & 1200 &  - & 0.38 & F 2.5 & 14.0 \l -5.06 & 7.06 & -4.7 & 5.65 &  2.7 & 20 \\
   74   & WN7 (WNE-w)   &  56.2 & 0.7 & 1300 &  0 & 1.50 & S     & 13.0 \r -5.17 & 5.29 & -4.6 & 5.4  &  7.2 & 15 \\
   75   & WN6-s         &  63.1 & 0.6 & 2300 &  0 & 0.93 & S     & 13.0 \r -5.58 & 5.94 & -4.1 & 5.7  & 19.0 & 22 \\
   78   & WN7h          &  50.1 & 1.0 & 1385 & 11 & 0.47 & S     & 11.5 \r -6.82 & 16.7 & -4.2 & 6.2  &  2.6 & 44 \\
   82   & WN7(h)        &  56.2 & 0.7 & 1100 & 20 & 1.00 & S     & 15.5 \l -7.22 & 14.9 & -4.0 & 6.3  &  3.1 & 51 \\
   84   & WN7 (WNE-w)   &  50.1 & 0.9 & 1100 &  0 & 1.45 & S     & 13.3 \l -5.67 & 8.89 & -4.6 & 5.65 &  3.2 & 20 \\
   85   & WN6h-w (WNL)  &  50.1 & 1.1 & 1400 & 40 & 0.82 & C 3.5 & 14.1 \l -7.22 & 21.1 & -4.2 & 6.4  &  1.7 & 59 \\

\s 87   & WN7h          &  44.7 & 1.3 & 1400 & 40 & 1.70 & S     & 12.3 \r -6.68 & 18.8 & -4.6 & 6.1  &  1.4 & 38 \\
   89   & WN8h          &  39.8 & 1.4 & 1600 & 20 & 1.58 & S     & 12.3 \r -7.25 & 26.5 & -4.5 & 6.2  &  1.7 & 44 \\
   91   & WN7  (WNE-s)  &  70.8 & 0.4 & 1700 &  0 & 2.12 & S     & 13.2 \l -5.67 & 5.00 & -4.0 & 5.75 & 14.2 & 23 \\
   94   & WN5-w         &  56.2 & 0.9 & 1300 &  - & 1.49 & C 3.4 & 10.1 \l -4.45 & 6.67 & -4.7 & 5.6  &  3.2 & 19 \\
   100  & WN7  (WNE-s)  &  79.4 & 0.3 & 1600 &  0 & 1.50 & S     & 13.0 \l -5.67 & 3.97 & -4.1 & 5.75 & 12.6 & 23 \\
   105  & WN9h          &  35.5 & 1.1 &  800 & 17 & 2.15 & S     & 11.0 \r -6.90 & 21.1 & -4.5 & 5.8  &  2.2 & 25 \\
   107  & WN8           &  50.1 & 0.8 & 1200 &  - & 1.41 & C 3.7 & 14.6 \l -7.22 & 16.7 & -4.0 & 6.2  &  3.9 & 44 \\
   108  & WN9h          &  39.8 & 1.4 & 1170 & 27 & 1.00 & S     & 13.3 \l -7.22 & 25.1 & -4.6 & 6.15 &  1.0 & 41 \\
   110  & WN5-s         &  70.8 & 0.5 & 2300 &  0 & 0.90 & C 3.5 & 10.6 \l -4.45 & 3.15 & -4.3 & 5.35 & 23.2 & 14 \\
   115  & WN6-w         &  50.1 & 0.9 & 1280 &  0 & 1.50 & S     & 11.5 \r -5.33 & 8.89 & -4.5 & 5.65 &  4.3 & 20 \\
\hline 
\end{tabular}
\newline

(to be continued) 
\normalsize
\end{flushleft}
\end{table*}

\begin{table*}
\addtocounter{table}{-1}
\caption[]{(continued)} 
\small
\begin{flushleft}
\begin{tabular}{ l l r r r r r l r c r l r r }
\hline\hline 
\s WR & 
Spectral subtype &
$T_*$ & 
$\log R_{\rm t}$ & 
\multicolumn{1}{c}{$\varv_\infty$} & 
$X_{\rm H}$ & 
$E_{b-\varv}$ & 
Law\a  & 
\DM\ \hspace{5mm} $M_\varv$ &
$R_*$& log\,$\dot{M}^\dagger$ & 
log\,$L$ & 
$\frac{{\dot M} \varv_{\infty}}{L/c}$ & 
$M$ \\

&      
& 
[kK] & 
$[R_{\odot}]$ & 
\multicolumn{1}{c}{[km/s]} &
[\%] &
[mag]& 
$R_V$  & 
[mag]\ \ \ [mag] & 
$[R_{\odot}]$ & 
[$M_{\odot}$/yr] &
[$L_{\odot}$] &
&
[$M_{\odot}$] \\

   (1) & (2)           &(3)      & (4)   & \multicolumn{1}{c}{(5)}    
                                             &  (6) & (7)&  (8) & (9) \hspace{0.6cm} (10)
                                                                                   & (11) & (12) & (13)& (14) &(15) \\
\hline 
\s 116   & WN8h          & 39.8 & 0.8 &  800 & 10 & 1.75 & S     & 13.4 \l -7.22 & 21.1 & -4.0 & 6.0  &  3.9 & 33 \\
   120   & WN7 (WNE-w)   & 50.1 & 0.8 & 1225 &  0 & 1.25 & S     & 12.7 \r -5.52 & 8.39 & -4.4 & 5.6  &  5.7 & 19 \\
   123   & WN8 (WNE-w)   & 44.7 & 0.7 &  970 &  0 & 0.75 & C 2.8 & 15.7 \l -7.22 & 17.7 & -4.0 & 6.05 &  5.5 & 35 \\
   124   & WN8h          & 44.7 & 0.7 &  710 & 13 & 1.08 & C 2.9 & 14.6 \l -7.22 & 16.7 & -4.1 & 6.0  &  3.0 & 33 \\
   127   & WN3 + O9.5V   & \comp \\
   128   & WN4(h)-w      & 70.8 & 1.1 & 2050 & 16 & 0.32 & C 3.6 & 12.9 \l -3.84 & 3.54 & -5.2 & 5.45 &  2.2 & 16 \\
   129   & WN4-w         & 63.1 & 0.9 & 1320 &  0 & 0.85 & S     & 13.6 \l -3.84 & 3.75 & -5.1 & 5.3  &  2.8 & 13 \\
   130   & WN8(h)        & 44.7 & 1.0 & 1000 & 12 & 1.46 & S     & 13.8 \l -7.22 & 22.1 & -4.2 & 6.25 &  1.8 & 47 \\
   131   & WN7h          & 44.7 & 1.3 & 1400 & 20 & 1.15 & S     & 14.9 \l -7.22 & 23.7 & -4.4 & 6.3  &  1.3 & 51 \\
   133   & WN5 + O9I     & \comp \\

\s 134   & WN6-s         & 63.1 & 0.7 & 1700 &  0 & 0.47 & C 3.4 & 11.2 \r -5.07 & 5.29 & -4.4 & 5.6  &  7.8 & 19 \\
   136   & WN6(h)-s      & 70.8 & 0.5 & 1600 & 12 & 0.45 & S     & 10.5 \r -4.69 & 3.34 & -4.5 & 5.4  & 10.9 & 15 \\
   138   & WN5-w + B?    & \comp \\
   139   & WN5 + O6II-V  & \comp \\
   141   & WN5-w + O5V-III& \comp \\
   147   & WN8(h) + B0.5V& 39.8 & 0.9 & 1000 &  5 & 2.85 & S     & 10.4 \l -7.22 & 29.8 & -3.8 & 6.3  &  3.6 & 51 \\
   148\b & WN8h + B3IV/BH& 39.8 & 1.3 & 1000 & 15 & 0.83 & C 3.0 & 14.4 \l -7.22 & 26.5 & -4.5 & 6.2  &  1.0 & 44 \\
   149   & WN5-s         & 63.1 & 0.7 & 1300 &  0 & 1.42 & S     & 13.3 \l -4.45 & 3.97 & -4.7 & 5.35 &  5.2 & 14 \\
   151   & WN4 + O5V     & \comp \\
   152   & WN3(h)-w      & 79.4 & 1.1 & 2000 & 13 & 0.50 & C 3.2 & 12.2 \r -2.65 & 2.23 & -5.5 & 5.25 &  1.7 & 12 \\

\s 155   & WN6 + 09II-Ib & \comp \\
   156   & WN8h          & 39.8 & 1.1 &  660 & 27 & 1.22 & S     & 13.3 \l -7.22 & 23.7 & -4.5 & 6.1  &  0.9 & 38 \\
   157   & WN5-w (+B1II) & \comp \\
   158   & WN7h + Be?    & 44.7 & 1.2 &  900 & 30 & 1.08 & S     & 14.3 \l -7.22 & 25.1 & -4.5 & 6.35 &  0.7 & 54 \\
\hline 
\end{tabular}
\newline

\begin{itemize}
\item[\a]  Column\,(8) Applied reddening law: S = Seaton (\cite{Seaton1979}), C = Cardelli et al. (\cite{Cardelli+al1989}), 
F = Fitzpatrick (\cite{Fitzpatrick1999}); for the last two, the given number is the adopted $R_V$ 

\item[\b]  Binary system in which the non-WR component contributes more than 15\% of the flux in the visual

\item[$^\dagger$]  Mass-loss rates are for an adopted clumping factor of $D$ = 4
\end{itemize}

\normalsize
\end{flushleft}
\end{table*}


\begin{figure}[!b]
\centering
\epsfxsize=\columnwidth
\mbox{\epsffile{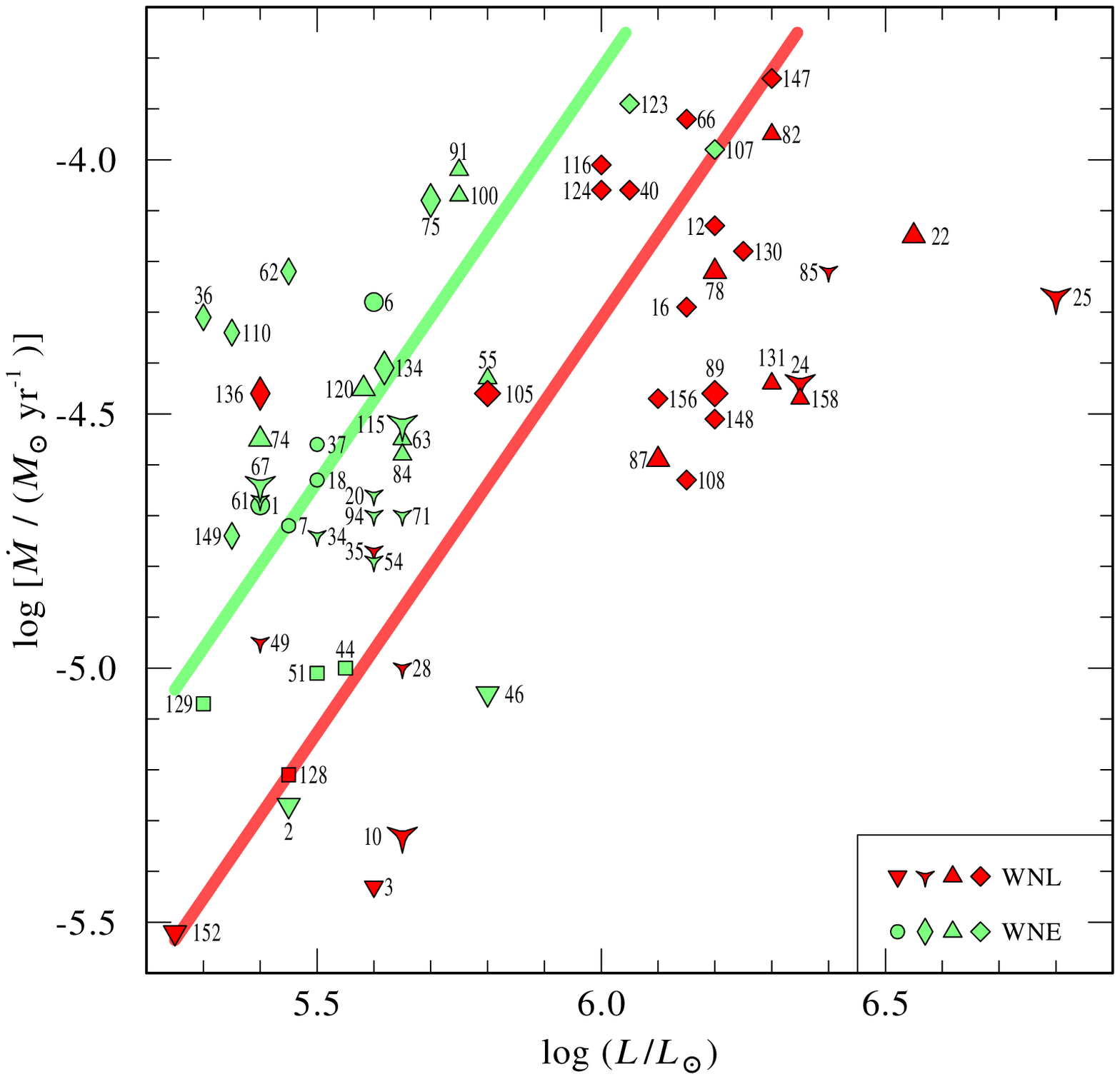}}
\caption{Empirical mass-loss rate versus luminosity for the Galactic WN
stars. Dark (red) filled symbols refer to stars with detectable
hydrogen (WNL), while light (green) filled symbols denote hydrogen-free
stars (WNE). The thick straight lines correspond to the mass loss --
luminosity relation proposed by Nugis \& Lamers
(\cite{NugisLamers2000}) for hydrogen-free WN stars (upper line) and
for a hydrogen mass fraction of 40\% (lower line).}
\label{fig:mdot}
\end{figure}

The most characteristic property of Wolf-Rayet stars is their strong
mass loss. Figure\,\ref{fig:mdot} shows the empirical mass-loss rate
obtained for our sample. The WR mass-loss rates are key ingredients in
various astrophysical models, e.g.\ of stellar evolution or galactic
evolution. In recent years, the empirical mass-loss formula from Nugis
\& Lamers (\cite{NugisLamers2000}) has become the most popular. We plot
their relation in Fig.\,\ref{fig:mdot} for comparison. The WNE stars
should be described by the upper line, while most WNL stars should lie
between that relation and the lower line, according to their hydrogen
abundance between zero and 40\% (mass fraction). For WNE stars, the
formula roughly gives the observed mass-loss rates. But the scatter is
large, and there is actually not much of a correlation between $\dot{M}$
and $L$. The same holds for the WNL stars. Even worse, most observed
mass-loss rates of the latter are significantly lower than described by
the Nugis \& Lamers formula. We conclude that more parameters (in
addition to $L$ and $X_{\rm H}$) are involved in controlling the mass
loss from WN stars.

Note that the empirical mass-loss rates scale inversely with the square 
root of the adopted clumping contrast, i.e.\ $\dot{M} \propto D^{-1/2}$. 
Our choice of $D$\,=\,4 is rather conservative. There are indications that 
the clumping is actually even stronger, and hence the mass-loss rates 
might still be over-estimated generally by a factor of 2 or 3.  

We also calculate the ratio between the wind momentum and the momentum
of the radiation field, $\eta = \dot{M} \varv_\infty c / L$
(column\,(14)). This number cannot exceed unity for a radiation-driven
wind if each photon can only be scattered (or absorbed) by {\em one}
spectral line. Models accounting for {\em multiple scattering} are not
bound to that limit. The first fully self-consistent hydrodynamical model
for a Wolf-Rayet star has been presented only recently by Gr\"afener \&
Hamann (\cite{wr111}). Their model for the WC5 prototype star WR\,111
exceeds the single-scattering limit moderately ($\eta$ = 2.54). WNL
models presently under construction (Gr\"afener \& Hamann, in
preparation) achieve ``efficiency numbers'' $\eta$ between 1 and 2,
typically. Our empirical values for $\eta$ are quite different for the
different subclasses; the arithmetic mean for the \WNEw\ and WNL
subclasses is 3.4 and 2.2, respectively. This is nearly reached by the
self-consistent models (the remaining difference might be attributed to
incomplete opacities). However, the mean $\eta$ of the \WNEs\ subclass
is 12.5, which is much higher than any of our radiation-driven models
can explain so far.

{\changed Remember that in Paper\,I we arrived at much higher values for
$\eta$ (9, 9, and 29 for the WNL, \WNEw, and WNEs\ subclass,
respectively). The smaller ``efficiency numbers'' result from the combined
effect of higher luminosities (due to line-blanketing) and lower
mass-loss rates (due to clumping). Even higher clumping contrast could
perhaps reduce $\eta$ by another factor of two.}

For completeness we derive a stellar mass from the luminosity 
by means of the mass-luminosity relation for helium stars from Langer 
(\cite{Langer1989}). Note that this relation might not be adequate for 
stars showing hydrogen. 

The discussion of the luminosities is deferred to 
Sect.\,\ref{sect:evolution} in the context of stellar evolution. First
we comment on the individual stars of our program, as far as they
deserve special remarks.

\begin{figure*}[!tb]
\centering
\epsfxsize=0.9\textwidth
\mbox{\epsffile{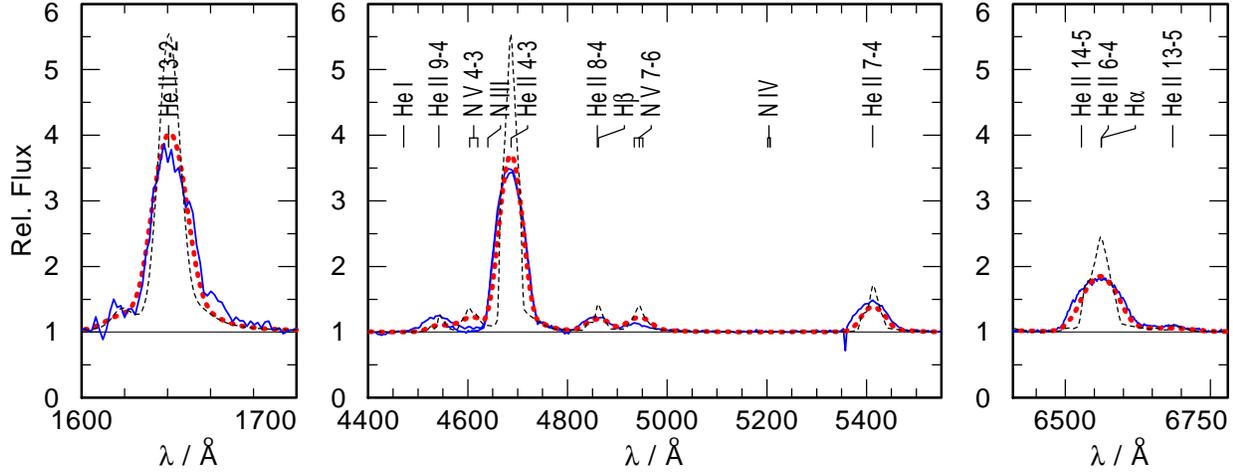}}
\caption{Observed line profiles of WR\,2 (solid line), compared to a 
model without rotation (dashed) and after convolution corresponding to 
$v_{\rm rot} \sin i$ = 1900 km/s (dotted).}
\label{fig:wr002}
\end{figure*}


\subsection{Comments on individual stars}

\label{sect:individual}

\noindent
{\bf WR\,1} belongs to the Cas\,OB7 association, for which the distance 
modulus has been revised from \DM\ = 12.1\,mag, as used in Paper\,I, 
to 11.3\,mag (Garmany \& Stencel \cite{Garmany+Stencel1992}).

\medskip\noindent
{\bf WR\,2} is marked as a visual binary in the WR catalog. But according
to Hipparcos, the companion is at 13.8\arcsec\ separation and should
therefore not contaminate the observed spectra.

A short glance at the atlas of Galactic WN spectra (Hamann et al.\
\cite{wn-atlas}) reveals that this weak-lined WN2 star is unique
with respect to the shape of its line profiles. Such round profiles are
not reproduced by any of our models, unless very rapid rotation is
assumed. Figure\,\ref{fig:wr002} compares observed  line profiles of
WR\,2 with model profiles before and after flux convolution (which is
only a very crude approach to a rotating wind, admittedly) with
1900\,km/s rotational velocity. Note that WR\,2, the only Galactic star
classified as WN2, is the hottest ($T_\ast = 140\,{\rm kK}$) and most
compact ($R_\ast = 0.89\,R_\odot$) star of our sample. A rotational
velocity of 1900\,km/s would imply that the star is spinning at its
break-up limit.

\medskip\noindent
{\bf WR\,3} is classified as WN3+O4 in the WR catalog, 
but there is not much evidence for its binary status.
Neither Massey \& Conti (\cite{Massey+Conti1981}) nor Marchenko et al.\
(\cite{Marchenko+al2004}) could confirm the previously claimed periodic
radial velocity variation. Our spectral fit gives no evidence of a
binary contamination. Thus we consider WR\,3 as a single star.

From our subtype calibration of $M_\varv$ we infer a distance of \DM\ = 
12.36\,mag, which is considerably smaller than the 13.17\,mag obtained by 
Arnal \& Roger (\cite{Arnal+Roger1997}) from studying the interstellar 
medium towards WR\,3. If their distance is right, WR\,3 would be 
unusually bright for its spectral subtype.   

In contrast to Paper\,I, we find that a model with about 20\% hydrogen
gives a slightly better fit of the H/He blends with H$\alpha$ and
H$\beta$ than do hydrogen-free models, supporting the corresponding result
of Marchenko et al.\ (\cite{Marchenko+al2004}). We therefore change the 
spectral type to WN3h-w. 

The stellar temperature that we obtain (89\,kK) is somewhat higher than 
the 77\,kK given by Marchenko et al.\ (\cite{Marchenko+al2004}). We also 
obtain a higher luminosity (by 0.2dex), due to the higher absolute 
visual magnitude that we adopt for this subtype.    

\medskip\noindent
{\bf WR\,6} shows considerable photometric, spectral, and polarimetric
variability with a period of about 3.7\,d; but this period is unstable, 
and there are also epochs without 
variations. Morel et al.\ (\cite{Morel+al.1997}) presented arguments for
why the variability cannot be attributed to the presence of a compact
companion, but is instead due to the rotation of a structured wind. Its
X-ray flux is strong for a single WN star, but too low for a high-mass
X-ray binary and within the usual $L_{\rm X}$-$L_{\rm bol}$-relation
for single O stars (Oskinova \cite{Oskinova2005}). The radio spectrum is
thermal (Dougherty \& Williams \cite{Dough2000}). Hence we consider WR\,6
as a single star.

\medskip\noindent
{\bf WR\,10} is designated as a visual binary, with spectral type WN5ha
(+A2V), in the WR catalog. However, the companion A star (for its spectrum, 
see Niemela et al.\ \cite{Niemela+al1999}) is at
3.7\arcsec\ separation and 1.7\,mag fainter (Hipparcos photometry). Therefore
a spectral contamination should be small even when the companion is
within the aperture of the instrument. Our models can fit the
SED and the line spectrum perfectly.

\medskip\noindent
{\bf WR\,12} is classified as WN8h+?, based on radial velocity variations
(Niemela \cite{Niemela1982}; Rauw et al.\ \cite{Rauw+al1996}) and the eclipse
lightcurve (Lamontagne et al.\ \cite{Lamontagne+al1996}) with a period of
23.9\,d. However, no lines from the companion are seen (SB1). As
our models reproduce the line spectrum and the SED (the latter with
anomalous reddening), we assume that the spectrum is not contaminated,
but the star might have evolved in a close binary system. 

As the membership in the cluster Bo7 is only ``possible'' (Lundstroem \&
Stenholm \cite{LS1984}), we instead assume the typical WN8 brightness for
that star and deduce the spectroscopic distance. Our reddening estimate
compares well with the average reddening of the cluster ($E_{b-\varv}$ =
0.74\,mag), but the cluster distance (\DM\ = 13.8\,mag) would result in
an 0.6\,mag fainter $M_\varv$ than the ``typical'' WN8 brightness from
our subtype calibration.

\medskip\noindent
{\bf WR\,16} shows substantial, but not periodic photometric variability
(Balona et al.\ \cite{Balona+al1989}), and polarimetric evidence of a
non-spherical outflow (Schulte-Ladbeck \cite{Schulte-Ladbeck1994}). The
thermal radio spectral index (Dougherty \& Williams \cite{Dough2000})
and the non-detection of X-rays (Oskinova \cite{Oskinova2005}) 
speak against binarity. WR\,16 has been analyzed in detail with
line-blanketed models by Herald et al.\ (\cite{Herald01}). We adopt
$\varv_\infty$ and $X_{\rm H}$ from that paper. Their analysis agrees
well with our present results, except that they assume a different
absolute visual magnitude for that spectral subtype.

\medskip\noindent
{\bf WR\,18} is considered by Lundstroem \& Stenholm (\cite{LS1984}) 
as a ``possible'' member of the Car\,OB1 
association (\DM\ = 12.55\,mag). However, with the typical brightness 
from our subtype calibration, the star appears much closer (\DM\ = 
11.5\,mag). Therefore we reject the membership. On the other hand, its 
reddening, $E_{b-\varv}$ = 0.75\,mag, is even higher than the typical 
reddening in Car\,OB1 (cf.\ the definite members WR\,22, WR\,24, and 
WR\,25).      

\medskip\noindent
{\bf WR\,21}, classified as WN5\,+\,O4-6, is a SB2 spectroscopic binary 
with a well-known period of 8.25\,d, showing weak O-star
absorption features moving in antiphase to the WR emission lines
(Niemela \& Moffat (\cite{Niemela+Moffat1982}). Lamontagne et al.\
(\cite{Lamontagne+al1996}) use the eclipse lightcurve to derive that 
84\% of the visual brightness must be attributed to the O star 
companion. Therefore we must omit that star from our present analysis. 

\medskip\noindent
{\bf WR\,22}, classified as WN7h+O9III-V in the WR catalog, is
definitely a SB2 binary with a period of $P = 80.4$\,d. The object has a
thermal radio spectrum (Dougherty \& Williams \cite{Dough2000}) and
strong X-ray emission (Oskinova \cite{Oskinova2005}). From the diluted O
star absorption lines visible in the composite spectrum, Rauw et al.\
(\cite{Rauw1996}) estimate that the O star companion contributes about
1/8 to the the flux in the visual. We ignore this contribution in our
spectral analysis. The stellar temperature that we obtain is about
10\,kK higher than from previous, un-blanketed analyses (Paper\,I, and
Crowther et al.\ \cite{CrowtherII}). We adopt the hydrogen abundance of
44\% by mass from the latter work, while our Paper\,I arrived at a
slightly lower value (40\%).

WR\,22 is a definite member of Car\,OB1 (Lundstroem \& Stenholm
(\cite{LS1984}). The reddening parameters from our analysis conform
closely with the anomalous reddening of that region. The distance of the
Car\,OB1 association has been increased considerably from \DM\ =
12.1\,mag as used in Paper\,I to 12.55\,mag (Massey \& Johnson
\cite{Massey+Johnson1993}). With this revised distance, WR\,22 becomes
very luminous. Simple application of the mass-luminosity relation for
helium stars yields a stellar mass of 74\,$M_\odot$, while from the
binary orbit the mass of the WR component has been determined to
72\,$M_\odot$ (Rauw et al.\ \cite{Rauw1996}) or 55\,$M_\odot$
(Schweickhardt et al.\ \cite{Schweickhardt1999}).

\medskip\noindent
{\bf WR\,24} is also a Car\,OB1 member, and the revised distance leads 
to a very high luminosity, as in the case of WR\,22.

\medskip\noindent
{\bf WR\,25} is classified as WN6h+O4f in the WR catalog, but its 
binary status remains debated. The photometric and
polarization variability is small and stochastic (Drissen et al.\
\cite{Drissen1992}). Raassen et al.\ (\cite{Raassen2003}) found no X-ray
variations over the history of X-ray astronomy. Only very recently, the
detection of radial velocity variations {\changed with a period of about 
200 days} 
has been reported (Gamen \& Gosset, private communication).

The line spectrum of WR\,25 can be perfectly matched by our single-star
model, including the absorption features.  However, the bright near-IR
photometry makes it difficult to fit the SED. As already stated by
Crowther et al.\ (\cite{CrowtherII}) this requires an anomalous
reddening law with large parameter $R_\varv$. Note that this is also the
case for the neighboring stars in Car\,OB1, WR\,22 and WR\,24, but less
extreme (see Table\,\ref{table:parameters}). WR\,25 is also affected by 
the upward revision of the distance to the Car\,OB1 association (see 
WR\,22). 

The stellar temperature we obtain is drastically higher (50\,kK) than
found in the previous, un-blanketed analyses (Paper\,I: 36\,kK, 
Crowther et al.\ \cite{CrowtherII}: 31\,kK). The consequence of the
increased distance, high (anomalous) reddening and the revised stellar
temperature is a luminosity of $\log L/L_\odot$ = 6.8, which would
advance WR\,25 to the most-luminous galactic WR star and to one of the
most-luminous stars of our whole Galaxy. If WR\,25 is indeed that
luminous, its paramount X-ray brightness is just in line with the usual
$L_{\rm X}$-$L_{\rm bol}$-relation for single O stars (Oskinova
\cite{Oskinova2005}). No conclusion can be drawn from the radio spectral
index of WR\,25 (Dougherty \& Williams \cite{Dough2000}).

Alternatively, the high near-IR flux could be attributed to a cool
companion. If a blackbody of 5\,kK is assumed to contribute to the near-IR, 
the SED can be fitted with much lower reddening, and the deduced stellar
luminosity is only $\log L/L_\odot$ = 5.8 (see Hamann \& Gr\"afener
\cite{H+G-canada}).  

As revealed by observations with the Spitzer Space Telescope, the mid-IR
($\lambda >$  10$\mu$m) spectrum of WR\,25 is dominated by strong
emission that cannot be attributed to the WR star (Barniske et al.\ 
\cite{Barniske+al2006}).

We decide to keep WR\,25 in our analyzed sample, as the visual spectrum
is not contaminated by a binary. However, reddening and derived 
luminosity are uncertain, and we leave the star out of the sample when
comparing it with tracks for single-star evolution.  

\medskip\noindent
{\bf WR\,28} has never been analyzed before. It is classified as
WN6(h)\,+\,OB? in the WR catalog, but the binary suspicion seems to be
based only on the weakness (``dilution'') of the emission lines. As our
models can reproduce the observed spectrum, we conclude that there is no
evidence of binarity. For the $M_\varv$ calibration, we put WR\,28 in
the hydrogen-free WNE6-w subclass. Its hydrogen content is low, and
adopting the high brightness of WNL stars would imply an implausible
luminosity and distance for this star.

\medskip\noindent
{\bf WR\,31}, classified as WN4\,+\,O8V  in the WR 
catalog, is an SB2 system (Gamen \& Niemela
\cite{Gamen+Niemela1999}) with a period of 4.8\,d, and it also has a visual
companion that is 2.5 {\sc hipparcos} magnitudes fainter at 0.6\arcsec\
separation. Lamontagne et al.\ (\cite{Lamontagne+al1996}) use
the eclipse lightcurve to derive that 72\% of the visual brightness must be
attributed to the O star companion. Therefore we must omit that star
from our present analysis.

\medskip\noindent
{\bf WR\,35} is classified as WN6h\,+\,OB? in the WR 
catalog, but the suspicion of binarity seems to be based only on 
the weakness (``dilution'') of the emission lines. As our models can 
reproduce the observed spectrum, we conclude that there is no evidence 
of binarity.  

\medskip\noindent
{\bf WR\,36} is classified as WN5-6\,+\,OB? in the WR
catalog, but the suspicion of binarity seems to be based only on 
the weakness (``dilution'') of the emission lines. As our models can 
reproduce the observed spectrum, we conclude that there is no evidence 
of binarity. The spectrum shows no \NIII\ lines, therefore we change 
the classification to WN5 in accordance with Smith et al.\ 
(\cite{Smith+al1996}). 

\medskip\noindent
{\bf WR\,40} shows a large variability in all spectral bands and in
polarimetry. Matthews \& Moffat (\cite{Matthews+Moffat1994}) conclude
that the apparently chaotic variations are in fact harmonics of two
periods. The variability might indicate the presence of a compact
companion, but the absence of X-ray emission (Oskinova
\cite{Oskinova2005}) speaks against such assumption. Herald et al.\
(\cite{Herald01}) have analyzed WR\,40 with line-blanketed models,
including the determination of various elemental abundances. We adopt
$\varv_\infty$ and $X_{\rm H}$ from that paper. The stellar temperature
they obtain is similar to our result. Their value for $\log L$ is lower
because they assume a lower absolute magnitude for that subtype than our
subtype calibration predicts.

\medskip\noindent
{\bf WR\,44} is classified as WN4\,+\,OB? in the WR
catalog, but the binary suspicion seems to be based only on
the weakness (``dilution'') of the emission lines. As our models can
reproduce the observed spectrum, we conclude that there is no evidence
of binarity.

\medskip\noindent
{\bf WR\,46} is classified as WN3p+OB? in the WR catalog. A radial
velocity period of 0.31\,d was found by Niemela et al.\
(\cite{Niemela+al1995}) and later revised by Marchenko et al.
(\cite{Marchenko+al2000}) to 0.329\,d. Both papers propose an evolved
binary with a neutron star companion and an accretion disk. Additionally,
Steiner \& Diaz (\cite{Steiner+Diaz1998}) found a photometric period of
7.46\,d. Veen et al.\ (\cite{Veen+al2002}) favor a non-radial pulsator
instead of the binary scenario. In X-rays WR\,46 is unusually bright for
a WNE star, but this agrees with the usual $L_{\rm X}$-$L_{\rm
bol}$-relation for single O stars (Oskinova \cite{Oskinova2005}). The
stellar spectrum does not appear to be composite. We consider WR\,46 as
a single star until the binary status is eventually confirmed.
$\varv_\infty$ = 2300\,km/s is kept from Paper\,I, because our fits
indicate that the 2450\,km/s from Crowther et al.\ (\cite{CrowtherIV})
is somewhat too high.

According to Tovmassian et al.\ (\cite{Tovmassian+al1996}), WR\,46 is a
member of the Cru\,OB4.0 association, and therefore the \DM\ was 
slightly revised from \DM\ = 13.0\,mag to 13.05\,mag.

\medskip\noindent
{\bf WR\,47}, classified as WN6\,+\,O5V  in the WR catalog, is an SB2
system (Gamen \& Niemela \cite{Gamen+Niemela1999})  with a period of
6.2\,d. Lamontagne et al.\ (\cite{Lamontagne+al1996}) derive from the
eclipse lightcurve that 61\% of the visual brightness must be attributed
to the O star companion. Therefore we must omit that star from our
present analysis.

\medskip\noindent
{\bf WR\,51} is classified as WN4\,+\,OB? in the WR
catalog, but the binary suspicion seems to be based only on
the weakness (``dilution'') of the emission lines. As our models can
reproduce the observed spectrum, we conclude that there is no evidence
of binarity.

The WR catalog lists WR\,51 as a possible member of the Anon Cen\,OB
association. However, Lundstroem \& Stenholm (\cite{LS1984}) already
noted that this star has much higher reddening. 
Our analysis yields $E_{b-\varv}$ = 1.4 while the average
reddening in Anon Cen OB is $E_{b-\varv}$ = 0.22 (Lundstroem \& Stenholm
\cite{LS1984}), thus confirming that WR\,51 is a background object.

\medskip\noindent
{\bf WR\,63} has not been analyzed before. It is classified as
WN7\,+\,OB in the WR catalog, but the binary suspicion seems to be based
only on the weakness (``dilution'') of the emission lines and the
presence of absorption features. As our models can reproduce the
observed spectrum, we conclude that there is no evidence of binarity.

\medskip\noindent
{\bf WR 66} is classified as WN8(h)\,+\,cc?  in the WR
catalog. From the observed variability, Antokhin et al.\
(\cite{Antokhin+al1995}) suggest the possibles existence of a compact
companion spiraling in towards the WR star. In contrast, Rauw et al.\
(\cite{Rauw1996}) favor the non-radial pulsation aspect as proposed for
many WN8. We conclude that the evidence of close binarity is insufficient. 
The star also has a visual companion, 1.05 {\sc hipparcos} magnitudes fainter
and separated by 0.4\arcsec.

Lundstroem \& Stenholm (\cite{LS1984}) and the WR catalog
list WR\,66 as a possible member of the association Anon Cir OB1. While
the reddening would be compatible, the distance module of Cir OB1
(12.57\,mag) would lead to an atypically low brightness for a WN8 subtype.
Therefore we instead consider WR\,66 as a background object.

\medskip\noindent
{\bf WR\,67} is classified as WN6\,+\,OB? in the WR catalog, but the
binary suspicion seems to be based only on the weakness (``dilution'')
of the emission lines. As our models can reproduce the observed
spectrum, we conclude that there is no evidence of binarity. WR\,67
belongs to the Cir\,OB1 association, for which the distance has been
slightly revised from \DM\ = 12.8\,mag to 12.6\,mag (Lortet et al.\
\cite{Lortet+al1987}).

\medskip\noindent
{\bf WR\,71}, classified as WN6\,+\,OB? in the WR catalog, is suspected
as a binary because of  ``diluted emission lines'' and because
Isserstedt et al.\ (\cite{Isserstedt+al1983}) found a photometric and
spectroscopic period of 7.69\,d. Balona et al.\ (\cite{Balona+al1989})
could confirm the photometric variability, but found large scatter
from a periodicity. Marchenko et al.\ (\cite{Marchenko+al1998b}) could
not even see any variability. We consider the evidences of binarity as 
not being sufficient. 
As we have no optical spectrum of WR\,71 at our disposal, the analysis
was based on the IUE data and no hydrogen abundance could be determined.

\medskip\noindent
{\bf WR\,78} is quoted here with a hydrogen abundance
of 11\% by mass from Crowther et al.\ (\cite{CrowtherII}), while 
our Paper\,I gives a slightly higher value (15\%).

\medskip\noindent
{\bf WR\,85} has not been analyzed before. It is classified as
WN6h\,+\,OB? in the WR catalog, but the binary suspicion seems to be
based only on the weakness (``dilution'') of the emission lines. As our
models can reproduce the observed spectrum, we conclude that there is no
evidence of binarity. The VB (visual binary) noted in the WR catalog
refers to a bright ($V$ = 6.5\,mag) G star that is 15\arcsec\ apart and
should not confuse the observed spectra.
 
\medskip\noindent
{\bf WR 87} is classified as WN7h\,+\,OB in the WR catalog, but the
binary suspicion seems to be based only on the weakness (``dilution'')
of the emission lines and spectral absorption features. As our models can 
reproduce the observed spectrum, we conclude that there is no evidence 
of binarity.  

\medskip\noindent
{\bf WR\,89} is classified as WN8h\,+\,OB in the WR catalog, but the
binary suspicion seems to be based only on the weakness (``dilution'')
of the emission lines and spectral absorption features. The radio
spectral index is thermal (Dougherty \& Williams \cite{Dough2000}). As
our models can reproduce the observed spectrum, we conclude that there
is no evidence of binarity. The visual companion (``VB'') mentioned in
the WR catalog is 10\arcsec\ away and too faint to confuse the
observation.

\medskip\noindent
{\bf WR\,94} is new in our sample. The observations available to us do not 
allow determination of the hydrogen abundance. 

\medskip\noindent
{\bf WR 105}, one of the two WN9h stars in our sample, shows a 
non-thermal radio spectral index (Dougherty \& Williams 
\cite{Dough2000}), but as there are no other indications of binarity we 
consider it as a single star.

\medskip\noindent
{\bf WR\,107} is new in our sample. The observations available to us do not 
allow determination of the hydrogen abundance. 

\medskip\noindent
{\bf WR\,108} is classified as WN9h+OB in the WR catalog. However, no
radial variations were found by Lamontagne et al.\
(\cite{Lamontagne+al1983}). The binary suspicion seems to be based only
on the weakness (``dilution'') of the emission lines and spectral
absorption features. As our spectral fit is perfectly able to reproduce
the SED and the line spectrum, we conclude that there is no evidence of
a luminous companion. The same conclusion was drawn by Crowther et al.\
(\cite{CrowtherI}), who presented a detailed analysis of that star.
However, their un-blanketed analysis (as well as ours in Paper\,I) gave
a significantly lower $T_\ast$ ($\approx$30\,kK) than the 40\,kK
obtained in the present paper. Consequently, we obtain a higher
luminosity. In agreement with Crowther et al.\ (\cite{CrowtherI}), we do
not consider WR\,108 as a member of Sgr\,OB1 (\DM\ = 11.0\,mag), but
assume the $M_\varv$ from our subtype calibration. $\varv_\infty$ and
$X_{\rm H}$ are taken from Crowther et al.\ (\cite{CrowtherI}).

\medskip\noindent
{\bf WR\,110} requires an anomalous reddening to fit the SED, which shows
relatively high flux in the near-IR. Its X-ray flux is strong for a
single WN star, but lies on the usual $L_{\rm X}$-$L_{\rm bol}$-relation
for single O stars (Oskinova \cite{Oskinova2005}). As there are no
further indications of binarity, we consider the star as single.
We classify this star as WN5 (WR catalog: WN5-6). 

This star is listed in Lundstroem \& Stenholm (\cite{LS1984}) as a
possible member of the Sgr\,OB1 association (\DM\ = 11.0\,mag). As we
generally do not adopt such ``possible'' memberships, we consider the
distance as unknown and employ the $M_\varv$ from our subtype
calibration instead. However, it turns out that the obtained distance
(\DM\ = 10.6\,mag) is similar to the Sgr\,OB1 value. For the color excess
$E_{b-\varv}$, we obtain 0.90\,mag, while Lundstroem \& Stenholm
(\cite{LS1984}) give 0.6\,mag as average for that region of the
association. Thus the question of the membership remains open.

\medskip\noindent
{\bf WR\,115} is classified as WN6\,+\,OB? in the WR catalog, but the
binary suspicion seems to be based only on the weakness (``dilution'')
of the emission lines. As our models can reproduce the observed
spectrum, we conclude that there is no evidence of binarity.
WR\,115 belongs to the Ser\,OB1 association, for which the distance has 
been slightly revised from \DM\ = 11.7\,mag to 11.5\,mag (Hillenbrand 
et al.\ \cite{Hillenbrand+al1993}).

\medskip\noindent
{\bf WR\,120} is a ``possible'' member of the open cluster Do\,33 
(Lundstroem \& Stenholm \cite{LS1984}). Adopting the distance of that 
cluster yields an absolute visual magnitude for that star, which is in 
nice agreement with our subtype calibration. Moreover, we find a 
color excess $E_{b-\varv}$ of 1.25\,mag, almost identical with the average 
cluster reddening (1.3\,mag, Lundstroem \& Stenholm \cite{LS1984}). Thus 
WR\,120 is most likely a true cluster member. 

\medskip\noindent
{\bf WR\,123} is a WN8 star with unclear binary status. Moffat \& Shara
(\cite{Moffat+Shara1986}) report a photometric variability and two
visual companions at 5\arcsec\ and 18\arcsec\ apart, which are too faint
to cause the observed variations. Therefore they discuss the
possibility of a compact companion. Alternatively, Marchenko et al.\
(\cite{Marchenko+al1998b}) and Marchenko \& Moffat
(\cite{Marchenko+Moffat1998}) attribute the variability to non-radial
pulsations. We consider WR\,123 as a single star.

\medskip\noindent
{\bf WR\,124} is of WN8h subtype and very similar in variability to
WR\,123. It also remains in our single-star sample as the evidence of
binarity is weak. The hydrogen abundance from Paper\,I has been confirmed 
by  Crowther et al.\ (\cite{CrowtherII}).

\medskip\noindent
{\bf WR\,127} is an SB2 binary system (WN3\,+\,O9.5V) with a period of
9.6\,d. Lamontagne et al.\ (\cite{Lamontagne+al1996}) use the eclipse
lightcurve to derive that 58\% of the visual brightness must be
attributed to the O star companion. Therefore we must omit that star
from our present analysis.

\medskip\noindent
{\bf WR\,128} is classified as WN4(h)+OB? in the WR catalog. Antokhin
and Cherepashchuk (\cite{Antokhin+Cherepashchuk1985}) found this star to
be photometric variable with $P$ = 3.871\,d and eclipses in the V band.
They concluded that a neutron star with an optical bright accretion disk
accompanies the WR star. We consider WR\,128 as a single star until a
compact companion has been confirmed. The hydrogen abundance and
$\varv_\infty$ in Table\,\ref{table:parameters} are taken from the 
analysis by Crowther et al.\ (\cite{CrowtherIV}).

\medskip\noindent
{\bf WR\,131} is classified as WN7h\,+\,OB in the WR catalog, but the
binary suspicion seems to be based only on the weakness (``dilution'')
of the emission lines and the presence of absorption features. As our
models can reproduce the observed spectrum, we conclude that there is no
evidence of binarity.

\medskip\noindent
{\bf WR\,133}, classified as  WN5\,+\,O9I, is a SB2 binary system 
with a period of 112\,d. According to Underhill \& Hill (\cite{Underhill+Hill1994}), 
the O star contributes between 11\% and 70\% to the total visual 
brightness. Because of this contamination we omit this star from the 
analyzed sample. 

\medskip\noindent
{\bf WR\,134} is a WN6 with reported spectral variability on a period of
2.25\,d. Morel et al.\ (\cite{Morel+al1999}) extensively discuss two
alternative origins, a compact companion or a rotationally modulated
anisotropic outflow. They conclude that the latter hypothesis leads to 
better consistency with the observations. The X-ray luminosity is rather
low, and the radio spectral index (cf.\ (Dougherty \& Williams
\cite{Dough2000}) is thermal. Therefore we decide to keep WR\,134 in our
sample of single stars.

WR\,134 belongs to the Cyg\,OB3 association, for which the distance has 
been revised from \DM\ = 11.6\,mag to 11.2\,mag (Garmany \& Stencel 
\cite{Garmany+Stencel1992}).

\medskip\noindent
{\bf WR 136} shows, like WR\,134, significant spectral variability.
Koenigsberger et al.\ (\cite{Koenigsberger+al1980}) found a period of
4.5\,d and concluded that the star might have a neutron star companion.
Antokhin \& Cherepashchuk (1985) revised the period to 4.57\,d, while
Robert et al.\ (\cite{Robert+al1989}) could not confirm the periodicity
from polarization measurements. The X-ray luminosity is rather low, and
the radio spectral index (cf.\ Dougherty \& Williams \cite{Dough2000})
is thermal. Therefore we decide to keep WR\,136 in our sample of single
stars.

WR\,136 belongs to the Cyg\,OB1 association, for which the distance 
has been drastically reduced by Garmany \& Stencel 
(\cite{Garmany+Stencel1992}) from \DM\ = 11.3\,mag to 10.5\,mag.

\medskip\noindent
{\bf WR\,138} is classified as WN5w\,+\,B? in the WR catalog. It is a
long-period SB2 (Lamontagne et al.\ \cite{Lamontagne+al1982}: 1763\,d;
Annuk (\cite{Annuk1990}: 1538\,d). According to Lamontagne et al.\
(\cite{Lamontagne+al1982}), both stars are of similar brightness. Our fits
can perfectly reproduce the spectral energy distribution from UV to IR.
However, the observed spectrum clearly shows absorption features at
\HeI\,4471 and \HeI\,5876 that are obviously due to the OB companion.
These features have equivalent widths as expected from a B0.5Ia star
alone. Thus we cannot exclude that the visual flux is strongly
contaminated by the companion. Therefore we exclude WR\,138 from our
sample of analyzed single-star spectra.

WR\,138 also shows short-term variability, which has been attributed to
a close compact companion (period 2.33\,d), making it a triple system
(Lamontagne et al.\ \cite{Lamontagne+al1982}). However, such line
profile variations are more likely due to the stellar wind dynamics. The
X-ray flux of WR\,138 is strong for a single WN star, but lies on the
usual $L_{\rm X}$-$L_{\rm bol}$-relation for single O stars (Oskinova
\cite{Oskinova2005}).

The visual companion mentioned in the WR catalog is at 0.865\arcsec\
separation and about 2.87 {\sc hipparcos} magnitudes fainter and should not
confuse the observations.

\begin{figure*}[!t]
\centering
\epsfxsize=12.0cm
\mbox{\epsffile{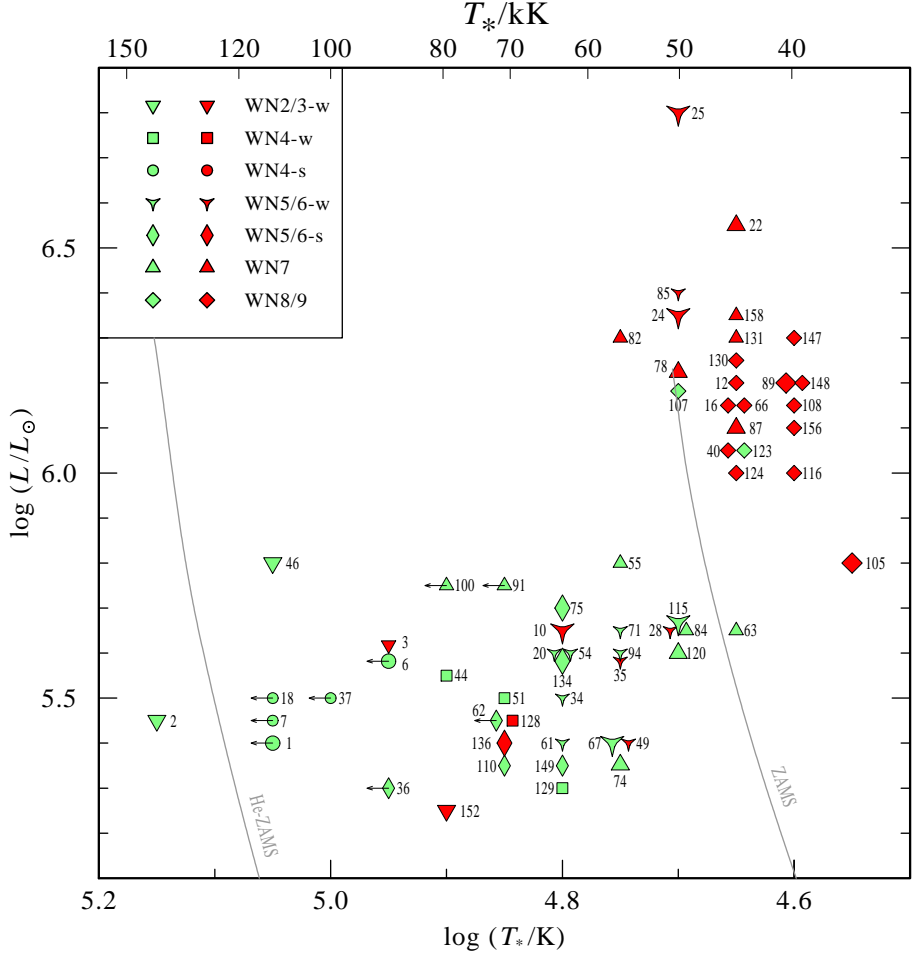}}
\caption{Hertzprung-Russell diagram of the Galactic WN stars.The bigger
symbols refer to stars whose distances are known from
cluster/association membership. The filling color reflects the surface
composition (dark/red: with hydrogen; light/green: hydrogen-free). The
symbol shapes are coding for the spectral subtype (see inlet). A little
arrow indicates that this particular star might be actually hotter
because of the parameter degeneracy discussed
in Sect.\,\ref{sect:linespectrum}}.
\label{fig:hrd}
\end{figure*}

\medskip\noindent
{\bf WR\,139}, better known as V444\,Cyg, is a WN5\,+\,O6III-V spectroscopic 
(SB2) and eclipsing binary with a period of 4.21\,d. The components are 
about equally bright (Hamann \& Schwarz \cite{Hamann+Schwarz1992}, 
Cherepashchuk et al.\ \cite{Cherepashchuk+al1995}), so we must 
exclude WR\,139 from our single-star analyses. 

\medskip\noindent
{\bf WR\,141} is classified as WN5w\,+\,O5V-III in the WR catalog. This
SB2 system has a period of 21.7\,d (Marchenko et al.\
\cite{Marchenko+al1998a}). The O star contributes 34\% to the total
visual brightness (Lamontagne et al.\ \cite{Lamontagne+al1996}).
Therefore we must exclude WR\,141 from our single-star analyses.

\medskip\noindent
{\bf WR\,147} is classified as WN8(h)\,+\,B0.5V in the WR catalog. This
extremely-long period binary has a period of 2880\,d and was spatially
resolved with HST by Lepine et al.\ (\cite{Lepine+al2001}), who estimated
the companion's spectral type as O5-7I-II(f). A non-thermal radio source
0.6\arcsec\ north of the the WR star is attributed to the colliding-wind
interaction zone (Watson et al.\ \cite{Watson+al2002}).

The WR star is 2.16\,mag brighter than the companion (Niemela et al.\
\cite{Niemela+al1998}), which means that the OB star contributes only
12\% to the visual flux. We neglect this contribution when analyzing the
spectrum. As the separation between the two components is too wide for a
strong evolutionary interaction, we also keep the star in our sample
when comparing it with single-star evolutionary tracks.

\medskip\noindent
{\bf WR\,148} is classified as WN8h\,+\,B3IV/BH in the WR catalog. The
radial velocity variations show a period of 4.317\,d (Moffat \&
Seggewiss \cite{Moffat+Seggewiss1980}, Drissen et al.\
\cite{Drissen+al1986}). Both prefer the interpretation that the
companion is a compact star or black hole. The large galactic height and
system velocity supports the scenario of a runaway after a supernova
kick. Panov et al.\ (2000) also found flare-like events and weak
X-ray emission, attributed to accretion processes on the companion. In
any case, additional light sources are 3\,mag fainter than the stellar
spectrum. Therefore we keep the star in our spectral analyses, but omit it
in our comparison with single-star evolutionary tracks.

\medskip\noindent
{\bf WR 151} is classified as WN4\,+\,O5V in the WR catalog. This SB2
system has a period of 2.1\,d, and the components are about equally
bright in the visual (Massey \& Conti \cite{Massey+Conti1981}).
Therefore we must omit this star from our spectral analysis.

\medskip\noindent
{\bf WR\,152} The hydrogen abundance and $\varv_\infty$ in
Table\,\ref{table:parameters} are taken from the analysis by Crowther et
al.\ (\cite{CrowtherIV}). This star belongs to the Cep\,OB1 association
for which the distance has been revised from \DM\ = 12.7\,mag to
12.2\,mag (Garmany \& Stencel \cite{Garmany+Stencel1992}).

\medskip\noindent
{\bf WR\,155}, also known as  CQ\,Cep, is classified as WN6\,+\,O9II-Ib
in the WR catalog. This double-lined (SB2) spectroscopic binary has a
period of 1.64\,d. Demicran et al.\ (\cite{Demicran1997}) found that
the components are about equally bright. Hamann \& Gr\"afener
(\cite{H+G-canada}) demonstrated that a slightly brighter companion is
needed to model the observed SED. Thus we must exclude the composite
spectrum of WR\,155 from our single-star analyses.

\medskip\noindent
{\bf WR\,156} is classified as WN8h+OB? in the WR catalog, obviously
because it is suspected to show ``diluted emission lines''. However, we
can match the spectrum with our models perfectly. Lamontagne et al.\
(\cite{Lamontagne+al1983}) found no radial velocity variations and
consider the photometric variability as intrinsic. We conclude
that this star is probably single. The hydrogen abundance (27\% by mass)
is taken from Crowther et al.\ (\cite{CrowtherII}), while our Paper\,I
gives a slightly higher value (30\%).

\medskip\noindent
{\bf WR\,157} is classified as WN5\,+\,B1II binary. Turner et al.\
(\cite{Turner+al1983}) found a visual B companion at 1 arcsec
separation, which seems to be the brighter component. Therefore the star
is omitted from our analyzed sample.

\medskip\noindent
{\bf WR\,158} is classified as WN7h\,+\,Be? in the WR catalog, because
Andrillat \& Vreux (\cite{Andrillat+Vreux1992}) found an
\ion{O}{i}\,8446 line indicative of Be emission line stars. However, they
also offer an alternative explanation by ``non-standard environmental
interaction''. Moreover, WR\,158 was suspected of being a binary because of
the weakness (``dilution'') of the emission lines. As our models can
reproduce the observed spectrum, we conclude that there is no
sufficient evidence of binarity.


\section{The evolution of massive stars} 
\label{sect:evolution}

Figure\,\ref{fig:hrd} shows the Hertzsprung-Russell diagram of our
program stars. The domains of the two spectral subclasses are clearly
separated.  Strikingly, the dividing line is just the hydrogen zero age
main sequence (ZAMS): while the WNL stars lie to the cooler side of the
ZAMS, the WNE stars populate a region between the hydrogen ZAMS and the
helium main sequence (He-ZAMS).

The WNL stars, generally showing a significant hydrogen abundance, have
stellar temperatures between 40 and 55\,kK and are very luminous (the
arithmetic mean of $\log L/L_\odot$ is 6.22 $\pm$ 0.20). Compared to
Paper\,I, this class has become significantly more luminous (by about
0.3\,dex in the mean $\log L$). The reason is a combination of effects:
higher stellar temperature, higher or anomalous reddening, larger
cluster distances. {\changed The subtype calibration alone has only 
limited influence; adopting $M_\varv$ = \mbox{-6.91\,mag} 
as the typical WNL brightness  
would shift the WNLs with unknown distance (small symbols in 
Fig.\,\ref{fig:hrd}) by 0.12dex to lower luminosities.}

The WNE stars, generally free of hydrogen, populate a range of lower
luminosities. Their average ($\log L/L_\odot$ = 5.5) is not revised
compared to Paper\,I. The range in $\log L$ became more narrow, as
several \WNEs\ stars that were considered to be very luminous in
Paper\,I, such as WR\,134 and WR\,136, became less luminous in the 
present paper because their cluster distances have been revised downward.

\begin{figure}[!b]
\centering
\epsfxsize=\columnwidth
\mbox{\epsffile{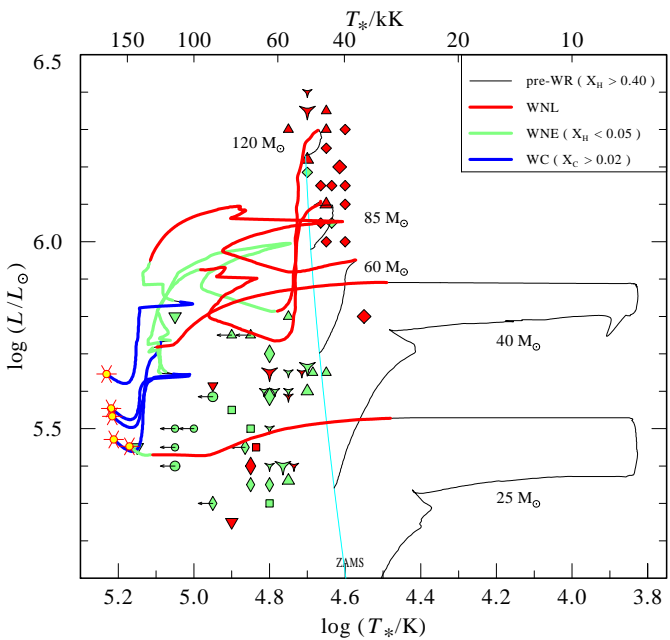}}
\caption{Hertzprung-Russell diagram with the analyzed WN stars,
now omitting the close binaries. 
The symbols have the same meaning as in Fig.\,\ref{fig:hrd}. The
evolutionary tracks from Meynet \& Maeder (\cite{MM03}) 
account for the effects of rotation (labels: initial mass). 
Thick lines refer to the different
WR phases (see inlet).} 
\label{fig:hrd+tracks}
\end{figure}

The main objective for analyzing WR stars is to understand the evolution 
of massive stars. At the time of Paper\,I, the most advanced stellar 
evolution models were from Schaller et al.\ (\cite{Schaller+al1992}). 
These tracks were not in good agreement with the empirical HRD of
Paper\,I.

\begin{figure}[!b]
\centering
\epsfxsize=\columnwidth
\mbox{\epsffile{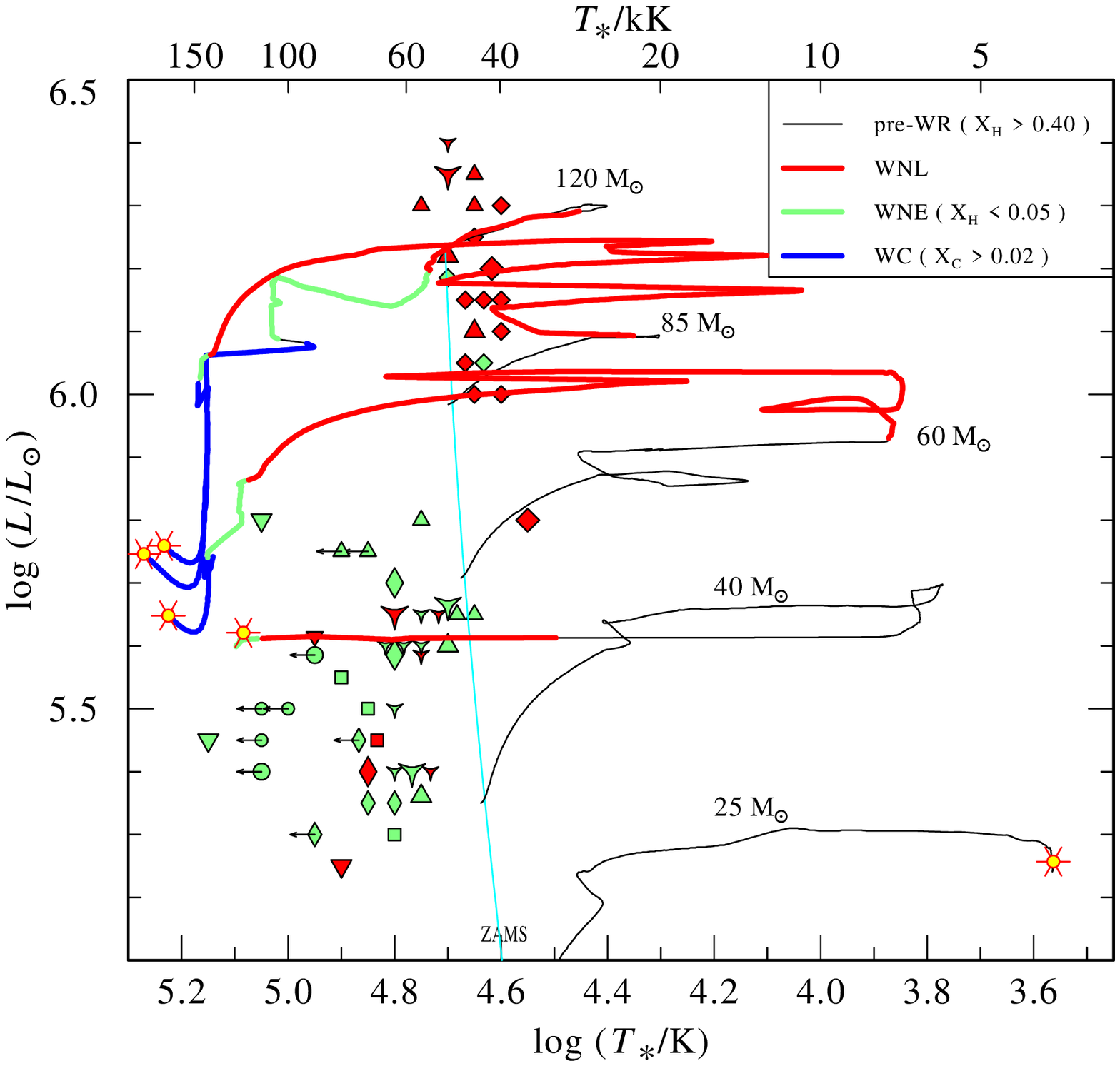}}
\caption{Hertzprung-Russell diagram with the analyzed WN stars,
now omitting the close binaries. 
The symbols have the same meaning as in Fig.\,\ref{fig:hrd}. This
set of evolutionary tracks from Meynet \& Maeder (\cite{MM03}) 
is calculated for {\em zero rotation} (labels: initial mass). 
Thick lines refer to the different WR phases (see inlet).} 
\label{fig:hrd+tracks-norot}
\end{figure}

The empirical HRD of the WN, as obtained with the upgraded analyses
presented here, is now the basis for re-assessing the question whether the
evolution of massive stars is theoretically understood. In the
meantime, since Paper\,I, the Geneva group has further improved the
evolutionary models with up-to-date physics and input data. Most
important, the new tracks now come in a version that accounts for the
effects of rotation (Meynet \& Maeder \cite{MM03}). In
Figs.\,\ref{fig:hrd+tracks} and \ref{fig:hrd+tracks-norot} we plot the
empirical HRD (now omitting the four confirmed close binaries) 
together with these evolutionary tracks for ``typical''
rotation (initial $\varv_{\rm rot}$ = 300\,km/s) and for zero rotation,
respectively. The metallicity is solar ($Z$ = 0.02). By using different
drawing styles, sections of the tracks are assigned to the spectral
classes according to their surface composition: the WNL stage is reached
when the star becomes hydrogen-deficient ($X_{\rm H} < 0.40$) and hotter
than 20\,kK; if hydrogen drops below $X_{\rm H} < 0.05$, the star
turns into a WNE type; the WC stage is reached when carbon appears at the
surface ($X_{\rm C} > 0.02$).

The minimum mass for a star to reach any WR phase, which is
37\,$M_\odot$ without rotation, is reduced by rotation to 22 $M_\odot$.
While the track for 25\,$M_\odot$ initial mass ends with the supernova
explosion as a red supergiant in case of no rotation, the corresponding track
returns to the blue when ``typical'' rotation is included in the
evolutionary model. Hence only the tracks with rotation can produce
stars with the low luminosities observed in the WNE subclass. However,
the WNE surface composition is only reached close to the helium main
sequence, while observed WNE stars are scattered over cooler
temperatures.

The HRD region where the WNL stars are observed is crossed by 
corresponding tracks for both versions, with and without rotation.
Very massive stars are predicted to skip the LBV excursions
when rotation is included. 

We want to recall that the evolutionary models we employ here do
not account for magnetic fields. A dynamo mechanism (``Tayler-Spruit
dynamo'') has been proposed for massive stars, although there is no
empirical proof so far for its existence. Maeder \& Meynet
(\cite{Maeder+Meynet2005}) expect significant effects of the predicted
field on the stellar evolution, but corresponding sets of tracks are
not yet available.

Comparing the empirical HRD just with tracks can be misleading, as
lifetimes in different parts of the tracks might be very different. Our
sample comprises almost all WN stars from the earlier (the 6th) catalog
of Galactic WR stars (van der Hucht et al.\ \cite{vdH81}). Those WN
stars that have been added later to the recent WR catalog are usually
highly reddened (and mainly discovered in the IR). Hence we can assume
that our analyzed sample of WN stars is roughly complete for a
well-defined part of our Galaxy, namely that part that can be observed
with little interstellar absorption. We do not support the speculation
by Massey (\cite{Massey2003}) that the number of WN stars is
underestimated.

\begin{figure*}[!tb]
\centering
\epsfxsize=\textwidth
\mbox{\epsffile{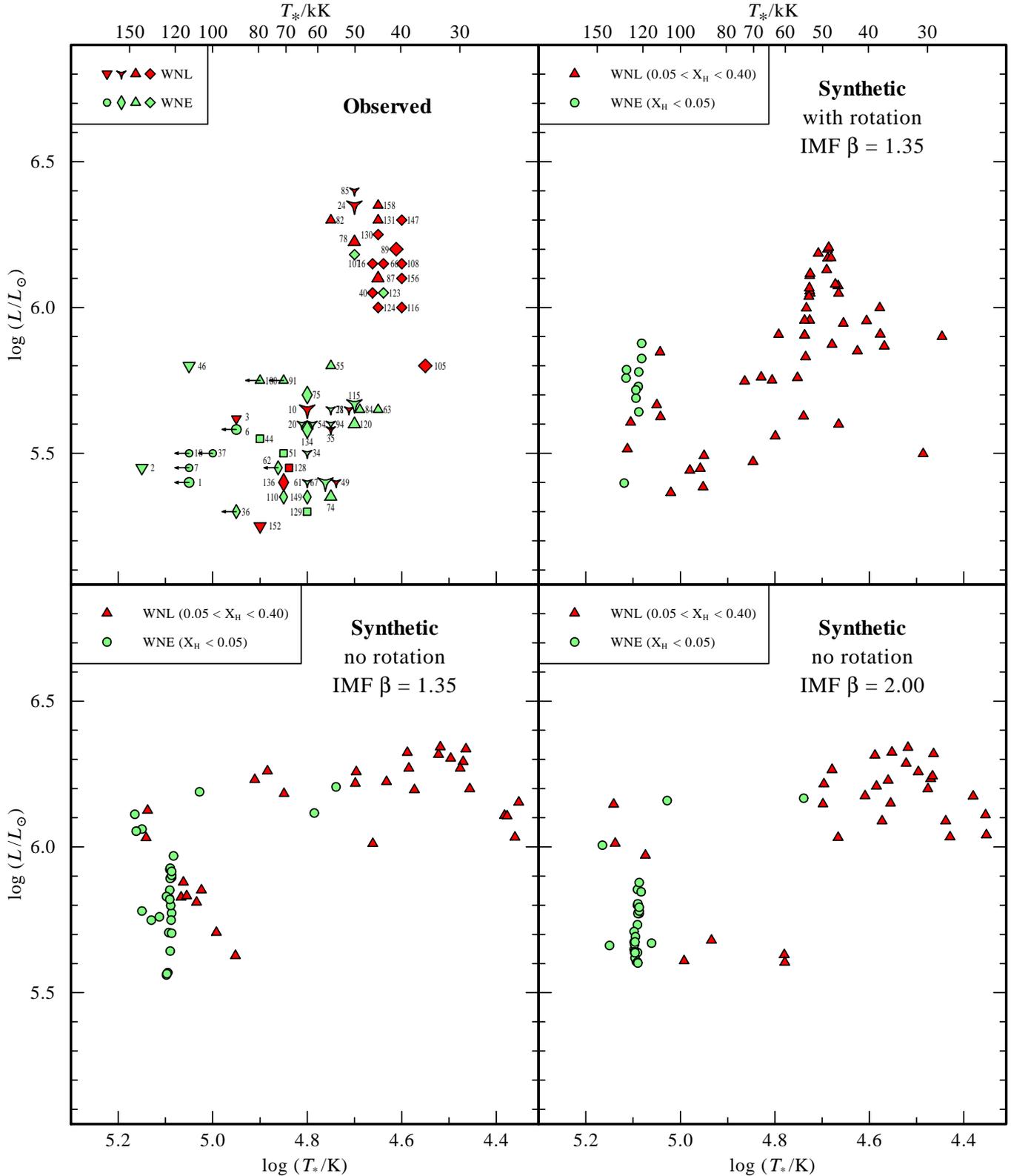}}
\caption{Comparison of the Galactic WN stars with synthetic
populations. {\em Top-left panel:} HRD of the analyzed Galactic WN 
stars (see Fig.\,\ref{fig:hrd} for the meaning of the symbols). Close
binaries are omitted. {\em Other panels:} Synthetic WN
population, based on the Geneva tracks (Meynet \& Maeder \cite{MM03}).
The filling color reflects the surface composition (dark/red: with
hydrogen; light/green: almost hydrogen-free). The synthetic
populations are for different assumptions as indicated in the plots. 
{\em Top-right panel:} 
evolution with rotation, Salpeter IMF ($\beta$ = 1.35);
{\em bottom-left panel:} 
evolution without rotation, Salpeter IMF ($\beta$ = 1.35); 
{\em bottom-right panel:} 
evolution without rotation and ``steep'' IMF 
($\beta$ = 2.00).} 
\label{fig:hrd+3syn}
\end{figure*}

Therefore we use the Geneva tracks to generate synthetic populations,
which we can compare with the empirical HRD. We adopt the usual power
law for the initial mass function (IMF) with exponent $\beta$ ($dN =
M^{-\beta-1}\,dM$). The widely accepted value for this exponent is
$\beta$ = 1.35 (``Salpeter IMF''), but there are indications that for
massive stars in the Galactic field this exponent is somewhat higher,
$\beta \ga$ 1.8 (Kroupa \& Weidner \cite{Kroupa+Weidner2003}). The star
formation rate is set constant; this is only a rough approximation, as
the analyzed sample contains several groups of stars that belong to the
same association and are therefore probably coeval, which might not
average out over the whole sample.

Now age and initial mass of a star are randomly chosen. Unfortunately,
the  available tracks are widely spaced in mass and qualitatively too
different for a full interpolation. Therefore we apply each track as it
is for a corresponding mass bin (borders are at 22, 32, 55, 75, 100,
135 $M_\odot$; for the track without rotation, the border between the
first and the second bin is set at 37\,$M_\odot$). Only the luminosity is
scaled as $L \propto M_{\rm ini}^{1.2}$  within each mass bin. The
synthetic stars are finally assigned to a spectral class according to 
their surface composition, using the same criteria as described above
when we plotted the tracks in different styles.  

For the synthetic populations shown in Fig.\,\ref{fig:hrd+3syn} we
created as many WN stars as our analyzed sample comprises (i.e.\ 59,
without the proven close binaries). As the most conventional choice, we
first take the tracks with rotation and adopt a ``Salpeter IMF'' (i.e.\
$\beta$ = 1.35). The result, shown in the top-right panel of
Fig.\,\ref{fig:hrd+3syn}, is disappointing when compared to the
empirical HRD (top-left panel). Although the 25\,$M_\odot$ track
produces WN stars that are as low-luminous as observed for the WNE,
those synthetic low-luminosity WN stars are almost all of subtype WNL,
i.e.\ they show hydrogen at their surface. Their predicted $X_{\rm H}$
is even higher than for most of the WNL stars.

Note that the observed sample contains more WNE than WNL stars. Taking
the HRD positions, we find 39 WNE stars left of the ZAMS and 20 WNL
stars on the right. On the basis of the hydrogen abundances, we have 33 WNE
and 26 WNL. In contrast, the
number of WNE (i.e.\ nearly hydrogen-free) stars in the synthetic sample
is only 10. Hence the evolution predicts that less than one fifth of the
WN stars are WNE, while more than half of the WN stars are actually
observed to be hydrogen-free.

The stellar temperature of the WNE stars is another problem. In the 
evolutionary tracks the hydrogen-free surface appears only when the star 
has almost reached the helium main sequence. Observed WNE stars are 
mostly cooler, even when taking into account that some of the stars 
fall into the domain of parameter degeneracy. The 
other WNE stars have typically thin winds, i.e.\ the model continuum is 
formed at layers of low expansion velocity. The ``stellar temperature'' 
of such stars does not depend much on the definition of the reference 
radius; i.e.\ $T_{\rm eff}$ at $\tau$=2/3 is roughly equal to our 
$T_\ast$ at $\tau$ = 20, so we conclude that these stars have a
larger photospheric radius than predicted by the stellar models. A
speculative explanation is that there is a huge extended layer on top of
the ``real'' hydrostatic core that expands only slowly (subsonic
velocities), possibly driven by the ``hot iron bump'' in the mean
opacity.

The WNL group is also not reproduced well by the synthetic population,
because their typical luminosity is smaller than observed. No synthetic star
lies above $\log L/L_\odot$ = 6.2, while the empirical HRD shows eight
such stars of very high luminosity. Two of these stars have a known
distance (WR\,24 and WR\,78), while the others rely on the subtype
calibration. Even with this caveat, there seems to be a discrepancy. A
flatter IMF (e.g.\ $\beta$ = 0.35) would produce more WNL stars at
highest luminosities.

After finding so many discrepancies, one might wonder if the tracks {\em
without} rotation really lead to even larger disagreement with the
empirical HRD. In the lower-left panel of Fig.\,\ref{fig:hrd+3syn} we
display the corresponding synthetic population, again for the Salpeter
IMF. As expected, the low-luminous WNE stars are missing. Interestingly,
the luminous WNL group in this plot reproduces the high average
luminosity of the observed WNL sample nicely, and only the scatter in
temperature is larger. Remarkable is also the statistical distribution
between WNE and WNL, which is now 30:29, i.e.\ close the observed slight
WNE majority. With a steeper IMF, $\beta$ = 2.0, the dichotomy between
the WNE and WNL becomes even nicer, while their number ratio is not
affected. Only the temperature scatter of the WNE is of course also not
reproduced by the synthetic WNE, which are all sitting near the He-ZAMS.
Hence we must conclude that, based on the shown comparison between the 
analyzed Galactic WN sample and the synthetic populations, there is no 
preference for the tracks that account for the stellar rotation. 

One can discuss more statistical numbers that are predicted by the
evolutionary calculations. The ratio between the number of WC to WN
stars in the WR catalog is about 0.9, while the population synthesis 
yields only 17 WC stars with rotation and $\beta$ = 1.35, but 45 and 37 
WC stars without rotation and $\beta$ = 1.35 and 2.00, respectively 
(always compared to 59 WN stars). Thus the non-rotating models 
are again closer to the observation. 

However, Meynet and Maeder (\cite{MM03}) point 
out that the low WC-to-WN ratio predicted by the models with rotation 
fits better into the trend with metallicity when different galaxies are 
compared. {\changed On the other hand, Eldridge and Vink 
(\cite{Eldridge+Vink06}) show that this trend can also be explained 
without rotation effects, when the WR mass-loss rates depend on 
metallicity. Such dependence is theoretically predicted from theoretical 
models for radiation-driven WR winds (Vink \& de Koter 
\cite{Vink+deKoter05}, Gr\"afener \& Hamann \cite{Tartu06} and in 
prep.). Thus it seems that the WC/WN metallicity trend does not 
unambiguously support the evolutionary models with rotation.} 
  
The number of Galactic WR to O stars, about 0.1, is better reproduced 
by the rotating than by the non-rotating models (Meynet and Maeder 
\cite{MM03}). However, the question of incompleteness might be more 
severe for the number of O stars than for the WR stars. 

Summarizing the comparison between the results of our spectral analyses
of the Galactic WN stars and the predictions of the Geneva evolutionary
calculations, we conclude that there is rough qualitative agreement.
However, the quantitative discrepancies are still severe, and there is
no preference for the tracks that account for the effects of rotation.
We wonder how future evolutionary models that account for magnetic
fields will compare with our empirical HRD. Based on the presently
existing tracks, it seems that the evolution of massive stars is still
not satisfactorily understood.

\bigskip {\em Concluding remark.} In this paper we presented
quantitative spectral analyses of the Galactic WN stars, based on the
line-blanketed Potsdam Wolf-Rayet (PoWR) model atmospheres. We hope that
the obtained stellar and atmospheric parameters provide an empirical
basis for various studies about the origin, evolution, and physics of
the Wolf-Rayet stars and their powerful winds.

\acknowledgements 

We thank L.\,Oskinova for many useful discussions and careful reading of
the manuscript. {\changed We also acknowledge the helpful suggestions of 
the referee, J.\,Vink.}


\begin{thebibliography}{}

\bibitem[1992]{Andrillat+Vreux1992}
Andrillat, Y., Vreux, J.-M., 1992, A\&A, 253, 37

\bibitem[1990]{Annuk1990}
Annuk, K., 1990, Acta Astronomica, 40, 267

\bibitem[1985]{Antokhin+Cherepashchuk1985}
Antokhin, I.I., Cherapashchuk, A.M., 1985, Soviet Astronomy Let\-ters, 11, 355

\bibitem[1995]{Antokhin+al1995}
Antokhin, I.I., Bertrand, J.-F., Lamontagne, R., Moffat, A.F.J., 1995, 
IAUS 163, 62

\bibitem[1997]{Arnal+Roger1997}
Arnal, E.M., Roger, R.S., 1997, MNRAS, 285, 253

\bibitem[1989]{Balona+al1989}
Balona, L.A., Egan, J., Marang, F., 1989, MNRAS, 240, 103
 
\bibitem[2006]{Barniske+al2006}
Barniske, A., Oskinova, L., Hamann, W.-R., Gr\"afener, G., 2006, in: 
Stellar Evolution at Low Metallicity: Mass Loss, Explosions, Cosmology. 
ASP Conf. Series, (in press)

\bibitem[1989]{Cardelli+al1989}
Cardelli, J.A., Clayton, G.C., Mathis, J.S., 1989, ApJ, 345, 245

\bibitem[1995]{Cherepashchuk+al1995}
Cherepashchuk, A.M., Koenigsberger, G., Marchenko, S.V., Moffat, A.F.J., 
1995, A\&A, 293, 142

\bibitem[1995a]{CrowtherI}
Crowther, P.A., Hillier, D.J., Smith, L.J., 1995a, A\&A 293, 172 

\bibitem[1995b]{CrowtherII}
Crowther, P.A., Hillier, D.J., Smith, L.J., 1995b, A\&A 293, 403 

\bibitem[1995c]{CrowtherIII}
Crowther, P.A.,  Smith, L.J., Hillier, D.J., Schmutz, W., 1995c, A\&A 293, 427 

\bibitem[1995d]{CrowtherIV}
Crowther, P.A.,  Smith, L.J., Hillier, 1995d, A\&A 302, 457

\bibitem[1995e]{CrowtherV}
Crowther, P.A.,  Smith, L.J., Willis, A.J., 1995e, A\&A 304, 269 

\bibitem[1997]{Demicran1997}
Demicran, O., Ak, H., \"Ozdemir, S., et al., 1997, AN, 318, 267

\bibitem[1986]{Drissen+al1986}
Drissen, L., Lamontagne, R., Moffat, A.F.J., Bastien, P., Seguin, M., 
1996, ApJ, 304, 188

\bibitem[1992]{Drissen1992}
Drissen, L., Robert, C., Moffat, A.F.J., 1992, ApJ, 386, 288

\bibitem[2000]{Dough2000}
Dougherty, S.M., Williams, P.M., 2000, MNRAS, 319, 1005

{\changed
\bibitem[2006]{Eldridge+Vink06}
Eldridge J.J., Vink J.S., 2006, A\&A (in press)
}

\bibitem[1999]{Fitzpatrick1999}
Fitzpatrick, E.L., 1999, PASP, 111, 63

\bibitem[1999]{Gamen+Niemela1999}
Gamen, R., Niemela, V.S., 1999, Rev. Mex. Astron. Astrofis. Ser. de Conf., 8, 55

\bibitem[1992]{Garmany+Stencel1992}
Garmany, C.D., Stencel, R.E., 1992, A\&AS, 94, 211

\bibitem[2005]{wr111}
Gr\"afener, G., Hamann, W.-R. 2002, A\&A, 432, 633

{\changed
\bibitem[2006]{Tartu06}
Gr\"afener, G., Hamann, W.-R. 2006, in: Stellar Evolution at Low 
Metallicity: Mass Loss, Explosions, Cosmology. Eds.
H.J.G.L.M. Lamers, N. Langer, T. Nugis \& K. Annuk. ASP Conference Series 
(in press)
}

\bibitem[2002]{blanketing02}
Gr\"afener, G., Koesterke, L., Hamann, W.-R. 2002, A\&A, 387, 244

\bibitem[1992]{Hamann+Schwarz1992}
Hamann, W.-R., Schwarz, E., 1992, A\&A, 261, 523

\bibitem[2004]{blanket-WN}
Hamann, W.-R., Gr\"afener, G., 2004, A\&A, 427, 697 

\bibitem[1998]{HK98}
Hamann, W.-R., Koesterke, L., 1998, A\&A, 333, 251 (Paper\,I)

\bibitem[2006]{H+G-canada}
Hamann, W.-R., Gr\"afener, G., 2006, in: Massive Stars in Interacting
Bina\-ries. N. St-Louis \& A.F.J. Moffat (eds.), ASP Conf. Ser. (in press)

\bibitem[1995]{wn-atlas}
Hamann, W.-R., Koesterke, L., Wessolowski, U., 1995, A\&AS, 113, 459

\bibitem[2001]{Herald01}
Herald J.E., Hillier, D.J., Schulte-Ladbeck, R.E., 2001, ApJ, 548, 932

\bibitem[1993]{Hillenbrand+al1993}
Hillenbrand, L.A., Massey, P., Strom, S.E., Merrill, K.M., 1993, AJ, 
106, 1906

\bibitem[1998]{HiMi98}
Hillier, D.J., Miller, D.L.\ 1998, ApJ, 496, 407


\bibitem[1981]{vdH81}
van der Hucht K.A., Conti P.S., Lundstr\"om I., Stenholm B., 1981, Space
Sci. Rev. 28, 227

\bibitem[2001]{vdH}
van der Hucht, K.A., 2001, New Astronomy Reviews 45, 135

\bibitem[1983]{Isserstedt+al1983}
Isserstedt, J., Moffat, A.F.J., Niemela, V.S., 1983, A\&A, 126, 183

\bibitem[1980]{Koenigsberger+al1980}
Koenigsberger, G., Firmani, C., Bisiacchi, G.F., 1980, Rev. Mex. Astron. 
Astrofis. 5, 45

\bibitem[2003]{Kroupa+Weidner2003}
Kroupa, P., Weidner, C., 2003, ApJ, 598, 1076

\bibitem[1982]{Lamontagne+al1982}
Lamontagne, R., Moffat, A.F.J., Koenigsberger, G., Seggewiss, W., 1982, 
ApJ, 253, 230

\bibitem[1983]{Lamontagne+al1983}
Lamontagne, R., Moffat, A.F.J., Seggewiss, W., 1983, ApJ, 269, 596

\bibitem[1996]{Lamontagne+al1996}
Lamontagne, R., Moffat, A.F.J., Drissen L., et al., 1996, AJ, 112, 2227

\bibitem[1989]{Langer1989}
Langer, N., 1989, A\&A, 210, 93

\bibitem[2001]{Lepine+al2001}
Lepine, S.

\bibitem[1987]{Lortet+al1987}
Lortet, M.-C., Georgelin, Y.P., Georgelin, Y.M., 1987, A\&A, 180, 65

\bibitem[1984]{LS1984}
Lundstroem, I., Stenholm, B., 1984, A\&AS, 58, 163

\bibitem[2005]{Maeder+Meynet2005}
Maeder, A., Meynet, G., 2005, A\&A, 440, 1041

\bibitem[1998]{Marchenko+Moffat1998}
Marchenko, S.V., Moffat, A.F.J., 1998, ApJ, 499, 195

\bibitem [1998a]{Marchenko+al1998a}
Marchenko, S.V., Moffat, A.F.J., Eenens, P.R.J., 1998, PASP, 110, 1416

\bibitem [1998b]{Marchenko+al1998b}
Marchenko, S.V., Moffat, A.F.J., van der Hucht, K.A., et al., 1998, 
A\&A, 331, 1022

\bibitem[2000]{Marchenko+al2000}
Marchenko, S.V., Arias, J., Barba, R., et al., 2000, AJ, 120, 2101

\bibitem[2004]{Marchenko+al2004}
Marchenko, S.V., Moffat, A.F.J., Crowther, P.A., et al., 2004, MNRAS, 353, 153 

\bibitem[2003]{Massey2003}
Massey, P., 2003, ARA\&A, 41, 15

\bibitem[1981]{Massey+Conti1981}
Massey, P., Conti, P.S., 1981, ApJ, 244, 173

\bibitem[1993]{Massey+Johnson1993}
Massey, P., Johnson, J., 1993, AJ, 105, 980

\bibitem[1994]{Matthews+Moffat1994}
Matthews, J.M., Moffat, A.F.J., 1994, A\&A, 283, 493

\bibitem[2003]{MM03}
Meynet, G., Maeder, A., 2003, A\&A, 404, 975

\bibitem[2001]{Moneti+al2001}
Moneti, A., Stolovy, S., Blommaert, J.A.D.L., Figer, D.F., Najarro,
F., 2001, A\&A, 366, 106

\bibitem[1980]{Moffat+Seggewiss1980}
Moffat, A.F.J., Seggewiss, W., 1980, A\&A, 86, 87

\bibitem[1986]{Moffat+Shara1986}
Moffat, A.F.J., Shara, M.M., 1986, AJ, 92, 952 

\bibitem[1997]{Morel+al.1997}
Morel, T., St-Louis, N., Marchenko, S.V., 1997, ApJ, 482, 470

\bibitem[1999]{Morel+al1999}
Morel, T., Marchenko, S.V., Eenens, P.R.J., et al., 1999, ApJ, 518, 428

\bibitem[1982]{Niemela1982}
Niemela, V.S., 1982, in: IUA Sym. 99, 299

\bibitem[1982]{Niemela+Moffat1982}
Niemela, V.S., Moffat, A.F.J., 1992, ApJ, 259, 213

\bibitem[1995]{Niemela+al1995}
Niemela, V.S., Barba, R.H., Shara, M.M., 1995, IAU Symp. 163, 245

\bibitem[1998]{Niemela+al1998}
Niemela, V.S., Shara, M.M., Wallace, D.J., et al., 1998, AJ, 115, 2047

\bibitem[1999]{Niemela+al1999}
Niemela, V.S., Gamen, R., Morrell, N.I., Ben\'itez, S.G., IAU Symp. 193, 26

\bibitem[2000]{NugisLamers2000}
Nugis, T., Lamers, H.J.G.L.M., 2000, A\&A, 360, 227

\bibitem[2005]{Oskinova2005}
Oskinova, L.M., 2005, MNRAS, 361, 679

\bibitem[2003]{Raassen2003}
Raassen, A.J.J., van der Hucht, K.A., Mewe, R., et al., 2003, A\&A, 402, 653

\bibitem[1996]{Rauw+al1996}
Rauw, G., Vreux, J.-M., Gosset, E., Manfroid, J., Niemela, V.S., 1996, 
in: Wolf-Rayet stars in the framework of stellar evolution. Li\'ege, p. 303

\bibitem[1996]{Rauw1996}
Rauw, G., Vreux, J.-M., Gosset, E., Hutsem\'ekers, D., Magain, P., 
Rochowicz, K., 1996, A\&A, 306, 771

\bibitem[1989]{Robert+al1989}
Robert, C., Moffat, A.F.J., Bastien, P., et al., 1989, ApJ, 347, 1034

\bibitem[1992]{Schaller+al1992}
Schaller, G., Schaerer, D., Meynet, G., Maeder, A., 1992, A\&AS, 96, 269

\bibitem[1994]{Schulte-Ladbeck1994}
Schulte-Ladbeck, R.E., 1994, Ap\&SS, 221, 347

\bibitem[1999]{Schweickhardt1999}
Schweickhardt, J., Schmutz, W., Stahl, O., Szeifert, Th., Wolf, B., 1999, 
A\&A, 347, 127

\bibitem[1979]{Seaton1979}
Seaton, M.J., 1979, MNRAS, 187, 73

\bibitem[1968]{Smith1968}
Smith, L.F., 1968, MNRAS, 140, 409

\bibitem[1996]{Smith+al1996}
Smith, L.F., Shara, M.M., Moffat, A.F.J., 1996, MNRAS, 281, 163

\bibitem[1998]{Steiner+Diaz1998}
Steiner, J.E., Diaz, M.P., 1998, PASP, 110, 276

\bibitem[1988]{Steiner+al1988}
Steiner, J.E., Cieslinski D., Jablonski F.J., 1988, ASPC, 1, 67

\bibitem[1999]{Steiner+al1999}
Steiner, J.E., Cieslinski, D., Jablonski, F.J., Williams, R.E., 1999,
A\&A, 351, 1021

\bibitem[1996]{Tovmassian+al1996}
Tovmassian, H.M., Navarro, S.G., Cardona, O., 1996, AJ, 111, 306

\bibitem[1983]{Turner+al1983}
Turner, D.G., Moffat, A.F.J., Lamontagne, R., Maitzen, H.M., 1983, AJ, 88, 1199

\bibitem[1994]{Underhill+Hill1994}
Underhill, A.B., Hill, G.M., 1994. ApJ, 432, 770

\bibitem[2002]{Veen+al2002}
Veen, P.M., van Genderen, A.M., van der Hucht, K.A., 2002, A\&A, 385, 619

{\changed
\bibitem[2005]{Vink+deKoter05}
Vink J.S., de Koter A., 2005, A\&A 442, 587
}

\bibitem[2002]{Watson+al2002}
Watson, S.K., Davis, R.J., Williams, P.M., Bode, M.F., 2002, MNRAS, 334, 631

\end{thebibliography}
\end{document}